%% file: main.tex
\documentclass[conference]{IEEEtran}

\usepackage{tikz}
\usepackage{amsmath}
\usepackage{pgfplots}
\usepackage{graphicx}
\usepackage{subcaption}
\usepackage{float}
\usepackage{cuted, caption, makecell}
\usepackage{mathtools}
\usepackage{booktabs}

\usepackage{ulem}    %
\newcommand{\ndssparagraph}[1]{\noindent\textbf{#1:}}

\newcommand\blfootnote[1]{%
  \begingroup
  \renewcommand\thefootnote{}\footnote{#1}%
  \addtocounter{footnote}{-1}%
  \endgroup
}

\usepgfplotslibrary{fillbetween}

\definecolor{wrongultramarine}{rgb}{0.07, 0.04, 0.56}

\usepackage{breakurl}           %
\usepackage[hyphens]{url} 
\usepackage[hyperfootnotes=false]{hyperref}

\begin{document}
\bstctlcite{IEEEexample:BSTcontrol}

\title{\Large \textsc{Brokenwire} : \bf Wireless Disruption of CCS Electric Vehicle Charging} %

\author{
{\rm Sebastian Köhler\textsuperscript{\textdagger}}\\
University of Oxford
\and
{\rm Richard Baker\textsuperscript{\textdagger}}\\
University of Oxford
\and
{\rm Martin Strohmeier}\\
Armasuisse S+T
\and
{\rm Ivan Martinovic}\\
University of Oxford
}

\IEEEoverridecommandlockouts
\makeatletter\def\@IEEEpubidpullup{6.5\baselineskip}\makeatother
\IEEEpubid{\parbox{\columnwidth}{
    Network and Distributed System Security (NDSS) Symposium 2023\\
    27 February - 3 March 2023, San Diego, CA, USA\\
    ISBN 1-891562-83-5\\
    https://dx.doi.org/10.14722/ndss.2023.23251\\
    www.ndss-symposium.org
}
\hspace{\columnsep}\makebox[\columnwidth]{}}

\maketitle

\input{00-Abstract/abstract}

\input{01-Introduction/introduction}

\input{03-Background/background}

\input{04-ThreatModel/threatmodel}

\input{05-Attack/attack}

\input{06-Evaluation/evaluation}

\input{07-Discussion/impact}

\input{07-Discussion/discussion}

\input{0X-Countermeasures/countermeasures}

\input{0X-RelatedWork/relatedwork}

\input{0X-Conclusion/conclusion}

\bibliographystyle{IEEEtranS}
\bibliography{references}

\appendix
\input{0X-Appendices/appendices}

\end{document}

%% file: 00-Abstract/abstract.tex
\begin{abstract}

We present a novel attack against the Combined Charging System, one of the most widely used DC rapid charging technologies for electric vehicles (EVs).
Our attack, \textsc{Brokenwire}, interrupts necessary control communication between the vehicle and charger, causing charging sessions to abort. 
The attack requires only temporary physical proximity and can be conducted wirelessly from a distance, allowing individual vehicles or entire fleets to be disrupted stealthily and simultaneously.  
In addition, it can be mounted with off-the-shelf radio hardware and minimal technical knowledge. 
By exploiting CSMA/CA behavior, only a very weak signal needs to be induced into the victim to disrupt communication --- exceeding the effectiveness of broadband noise jamming by three orders of magnitude. 
The exploited behavior is a required part of the HomePlug Green PHY, DIN 70121 \& ISO 15118 standards and all known implementations exhibit it.

We first study the attack in a controlled testbed and then demonstrate it against eight vehicles and 20 chargers in real deployments. 
We find the attack to be successful in the real world, at ranges up to 47 m, for a power budget of less than 1 W. 
We further show that the attack can work between the floors of a building (e.g., multi-story parking), through perimeter fences, and from `drive-by' attacks. 
We present a heuristic model to estimate the number of vehicles that can be attacked simultaneously for a given output power.

\textsc{Brokenwire} has immediate implications for a substantial proportion of the around 12 million battery EVs on the roads worldwide --- and profound effects on the new wave of electrification for vehicle fleets, both for private enterprise and crucial public services, as well as electric buses, trucks and small ships.
As such, we conducted a disclosure to the industry and discussed a range of mitigation techniques that could be deployed to limit the impact.

\end{abstract}

%% file: 01-Introduction/introduction.tex
\section{Introduction}

Replacing\blfootnote{\textsuperscript{\textdagger} Both authors contributed equally to this paper.} petrol and diesel vehicles with electric vehicles (EVs) has been one of the main approaches to cut down the global greenhouse-gas emission.
As a result, many countries have committed to completely ban the sale of vehicles with combustion engines within the next decade~\cite{bbcban2030, independengermanyban, listofcountriesban}.
The US government announced the transition of their 645,000 vehicles to a fully electric fleet~\cite{biden_us_fleet}.
The National Health Service in the UK plans to purchase fully electric ambulances~\cite{nhsev} and the Ministry of Defence announced the introduction of fully electric battlefield vehicles~\cite{mod_evs}.
In addition to governmental institutions, delivery companies are moving towards EVs, too.
Amazon started switching to electric delivery vans~\cite{amazonev} and electric trucks~\cite{amazontrucks}.
The United States Postal Service (USPS), Royal Mail, United Parcel Service (UPS), and DPD announced the transition to fully electric delivery vans~\cite{ups_arrival, usps_ev, royal_mail_ev, dpd_ev}.

\begin{figure}[t]
  \centering
  \includegraphics[width=1\linewidth]{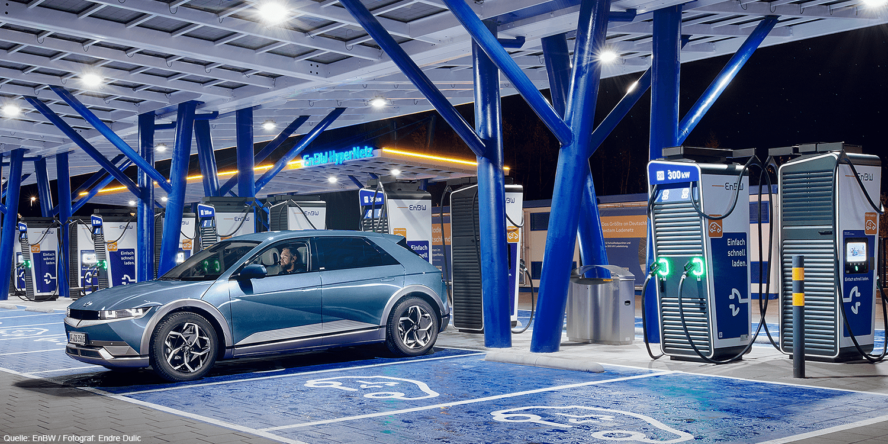}
  \centering
  \caption{Europe's largest high-power charging hub with 26 CCS charging stations that allow a total of 52 vehicles to be charged concurrently~\cite{enbw_kamen}.}
  \label{fig:hpc_hub} 
\end{figure}

\begin{figure*}[t]
	\centering
	\begin{subfigure}[b]{.33\linewidth}
		\centering
		\includegraphics[width=.98\textwidth]{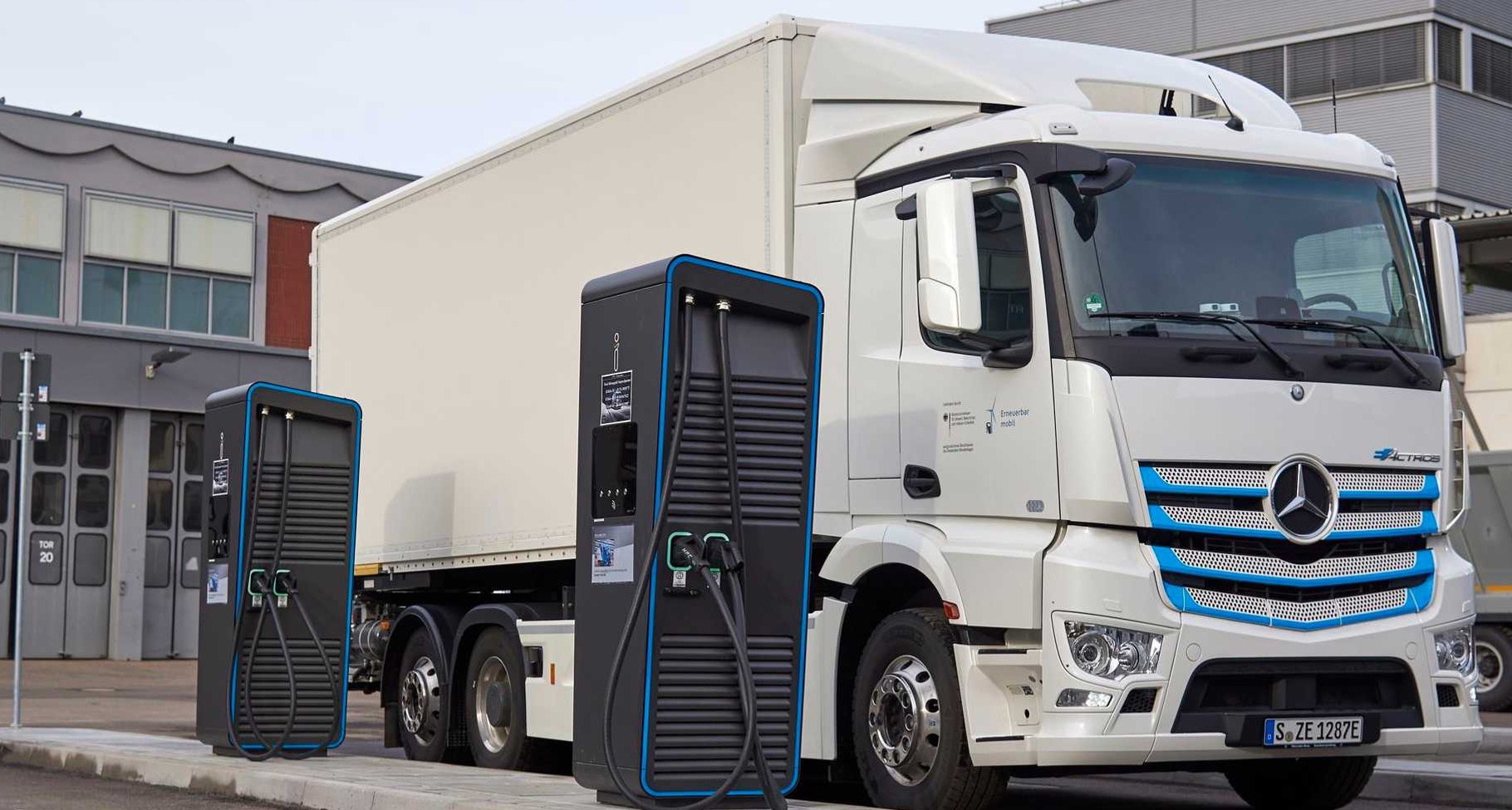}
		\caption{Heavy-duty truck using CCS~\cite{daimler_etruck}.}
		\label{fig:example-truck} 
	\end{subfigure}%
	\begin{subfigure}[b]{.33\linewidth}
		\centering
		\includegraphics[width=.98\textwidth]{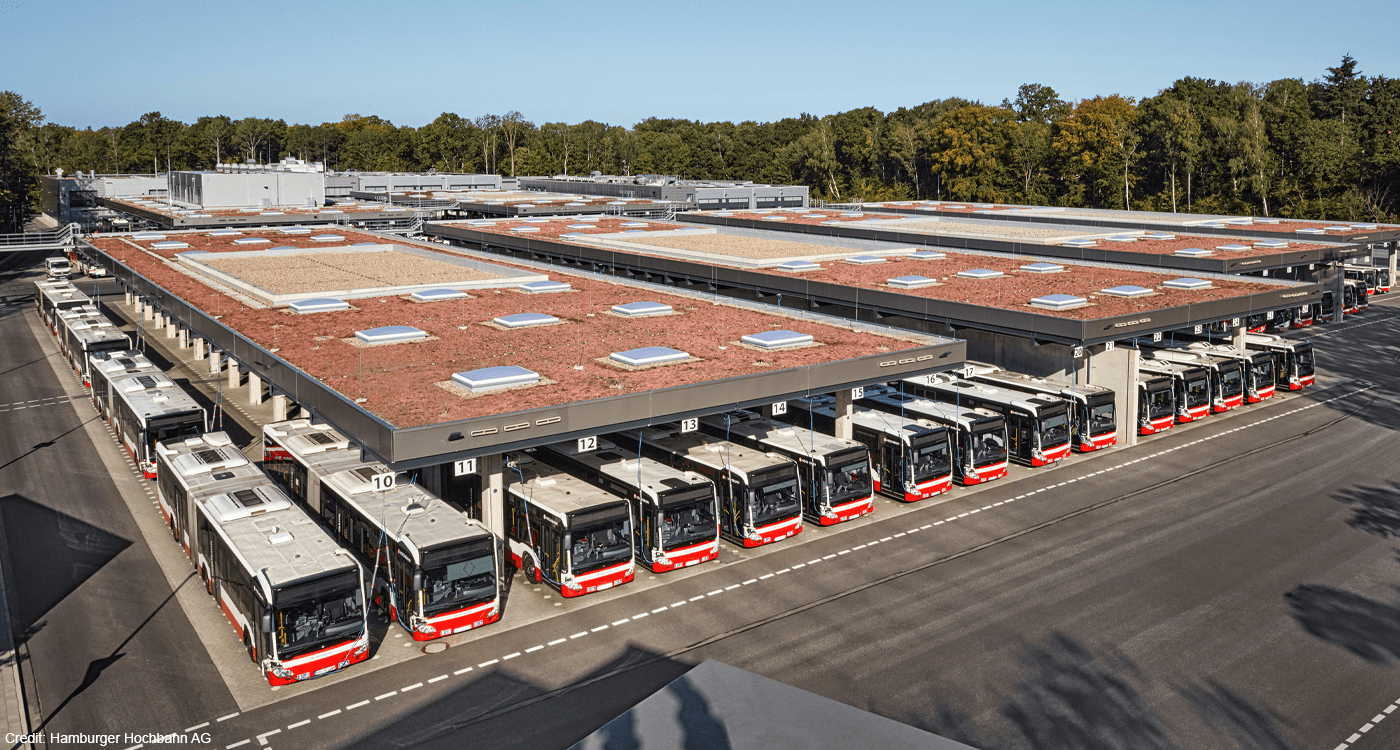}
		\caption{Bus depot with 46 CCS plugs~\cite{siemenshamburg}.}
		\label{fig:example-bus} 
	\end{subfigure}%
	\begin{subfigure}[b]{.33\linewidth}
		\centering
		\includegraphics[width=.98\textwidth]{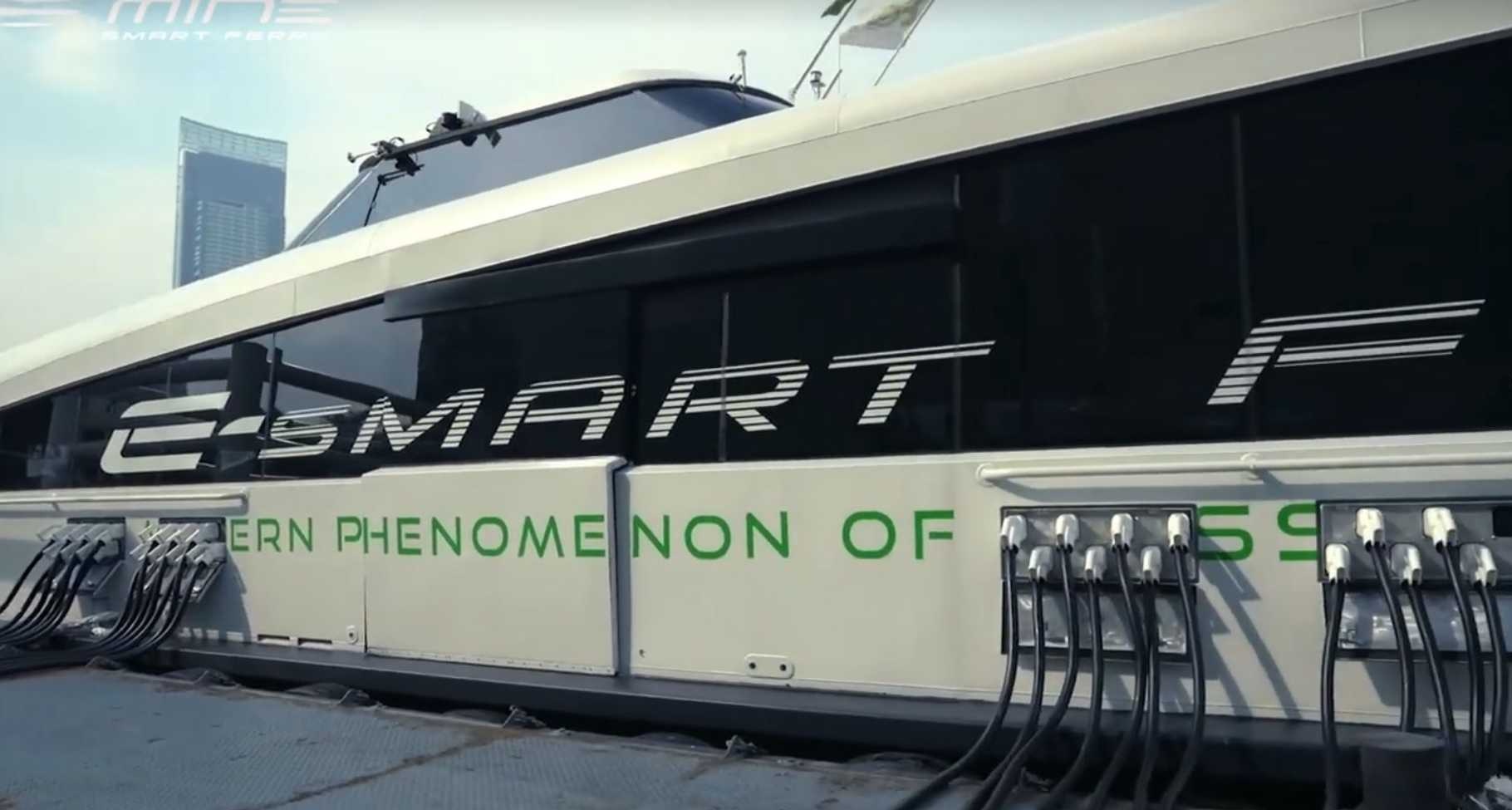}
		\caption{Ferry utilizing 28 CCS feeds~\cite{ferry_ccs_plugs}.}
		\label{fig:example-ferry} 
	\end{subfigure}%
	\caption{Examples of current electric vehicles now following the Combined Charging System standard and implementing the high-level communication using PLC, which makes them vulnerable to the \textsc{Brokenwire} attack.}
	\label{fig:examples-ccs}
\end{figure*}

Nevertheless, one disadvantage of EVs is that they are slower to refuel than fossil-fuel vehicles.
Therefore, the successful transition to all-electric vehicles requires a comprehensive and harmonized charging infrastructure that enables a vehicle to be charged in a short time~\cite{euStudy2018}.
This is achieved by increasing the charging capabilities of the charging stations, also known as Electric Vehicle Supply Equipment (EVSE), and building so-called charging hubs that enable the charging of multiple cars simultaneously.
Figure~\ref{fig:hpc_hub} shows such a charging hub, the largest currently in operation in Europe.

Direct Current (DC) charging has become the de-facto standard to enable rapid charging. 
For safety and efficiency reasons, the high-power DC charging stations rely on communication with the vehicle to exchange vital information, such as maximum voltage, required current, and the State of Charge (SoC). 
This means the availability of the communication link is crucial for the charging session and any disruption of the communication, intentional or unintentional, will result in the charging process being aborted for safety reasons~\cite{iso15118-3}.

The fastest-growing DC charging standard, already dominant across North America and Europe, is the Combined Charging System (CCS)~\cite{ev-sales-volumes,ev-charging-market-report}. 
CCS provides a high-bandwidth IP link via power-line communication (PLC) for the communication between the EV and EVSE. 
This brings benefits in terms of reusing commodity technologies and affording capacity for additional services, such as automatic billing and demand-response charging, in addition to the crucial charging session control~\cite{iso15118-1}. 

However, it is known that PLC, as used in CCS charging, unintentionally leaks communication signals via the charging cable~\cite{baker2019} and it has been shown in other settings that PLC is susceptible to electromagnetic interference~\cite{lampe2016}.

In this paper, we present \textsc{Brokenwire}, an attack that exploits the combination of the susceptibility of PLC to intentional electromagnetic interference (IEMI), the use of unshielded charging cables, and the application of a collision avoidance mechanism in the low-level communication protocol. 
We demonstrate that the charging communication required by CCS can be disrupted wirelessly --- causing the charging session to abort and leaving the vehicle and the EVSE in an error state, from which it will not automatically recover, even if the attack ceases. 
We show that only temporary physical proximity is required and that it is sufficient to execute the attack for only around two seconds --- just long enough to cause a timeout and for the charging to abort.
Therefore, the attack can reach beyond physical barriers and enables wardriving, which makes it stealthy, easily-deployed and challenging to detect.

Since CCS is becoming increasingly popular as a charging standard for a wide range of EVs --- beyond solely passenger vehicles --- the \textsc{Brokenwire} attack has immediate implications to a variety of other applications as well, such as emergency vehicles~\cite{nhsev}, buses~\cite{siemenshamburg}, heavy-duty trucks~\cite{ecascadia}, aircraft pushback tractors~\cite{pushbacktractor}, private boats~\cite{aqua_superpower}, public ferries~\cite{minesmartferry} and even airplanes~\cite{skycharger}. 
A handful of examples of vehicles that use CCS and are vulnerable to the \textsc{Brokenwire} attack are shown in Figure~\ref{fig:examples-ccs}. 
These illustrate both the range of affected vehicles and also the density with which chargers are arranged for fleet purposes and therefore their vulnerability to a ranged, wireless attack.
Moreover, CCS is also poised to play a decisive role in the future of the power grid~\cite{charinV2g2019}.
As the energy generated from renewable resources increases, the need for electricity storage has become more important than ever.
Bi-directional charging in combination with Vehicle-to-Grid (V2G) communication will enable vehicles to act as energy storage to buffer excess energy and feed it back into the grid to meet peak demand~\cite{mwasiluV2g2014}.
First trials of this approach have recently started in different countries, for example, Germany~\cite{bmw_group_v2g}, Switzerland~\cite{honda_v2x_switzerland} and the UK~\cite{bus2grid_london}.
As such, EVs with CCS will also soon play a significant role in the stability of the power grid, intertwining them even further into critical infrastructure.
Our work highlights a severe design flaw in the use of PLC for charging communication that leaves millions of vehicles, some of which are in constant use within critical infrastructure sectors, vulnerable. 

\noindent\textbf{Contributions}
\begin{itemize}
    \item We identify a vulnerability in the most widely adopted DC rapid charging standard in Europe and North America leaving millions of EVs vulnerable.
    \item We demonstrate the \textsc{Brokenwire} attack in an extensive evaluation, with both controlled laboratory and real-world environments.
    \item We analyze the effects of the \textsc{Brokenwire} attack and its real-world implications.
    \item We propose different countermeasures, including a cheap and easy-to-deploy hardware approach that makes the attack orders of magnitude harder to conduct.
\end{itemize}

%% file: 03-Background/background.tex
\section{Background} \label{sec:background}

This section describes the underlying technical concepts and terminologies of today’s charging standards, which help better understand the \textsc{Brokenwire} attack. 

\subsection{Electric Vehicle Charging} \label{sec:ev_charging_basics}

Charging an electric vehicle can be done with either Alternating Current (AC) or Direct Current (DC).
While for AC charging the car has to be equipped with a rectifier that converts the alternating current to direct current, for DC charging this process is outsourced to the charging station.
To reduce the additional weight from the rectifier, its capacity is limited, which caps the maximum charging power for AC charging to around 22 kW.
In contrast, DC charging enables high-power charging, often referred to as rapid charging, with up to 350 kW.
Therefore, for recharging a vehicle in a short time frame, DC chargers are the first choice.

In addition to the Combined Charging System, three competing DC charging standards currently exist --- CHAdeMO, Tesla's supercharger, and GB/T 20234.
While Tesla's supercharger is a proprietary technology, CHAdeMO was developed by car manufacturers from Japan and is an alternative to CCS mainly in the Asian market~\cite{das2020electric}.
Similarly, GB/T 20234 was designed for the Chinese market. 
The main difference between these standards and CCS is the use of CAN for the charging communication rather than PLC. 
However, more and more non-European car manufacturers previously using CHAdeMO, such as Kia, Nissan, and Hyundai, moved to CCS for markets outside of Asia~\cite{electrek2018, autocar2019, chargedevs2020}. 
Moreover, Tesla started using CCS with the introduction of the European version of their Model 3 in 2018~\cite{teslaCCS2018} and recently announced an updated version of their Model S and Model X for the European market, which will be equipped with a CCS socket by default and no longer require an adapter~\cite{models2022}.
The recent developments suggest that CCS will play a crucial role as a globally adapted standard in the future.
As such, our work focuses on CCS only and will discuss the relevant details about its underlying technical details in the subsequent section.

\subsection{Combined Charging System} \label{sec:ccs_basics}

\begin{figure}[t]
	\centering
	\begin{subfigure}[b]{.499\linewidth}
		\centering
		\includegraphics[width=.6\textwidth]{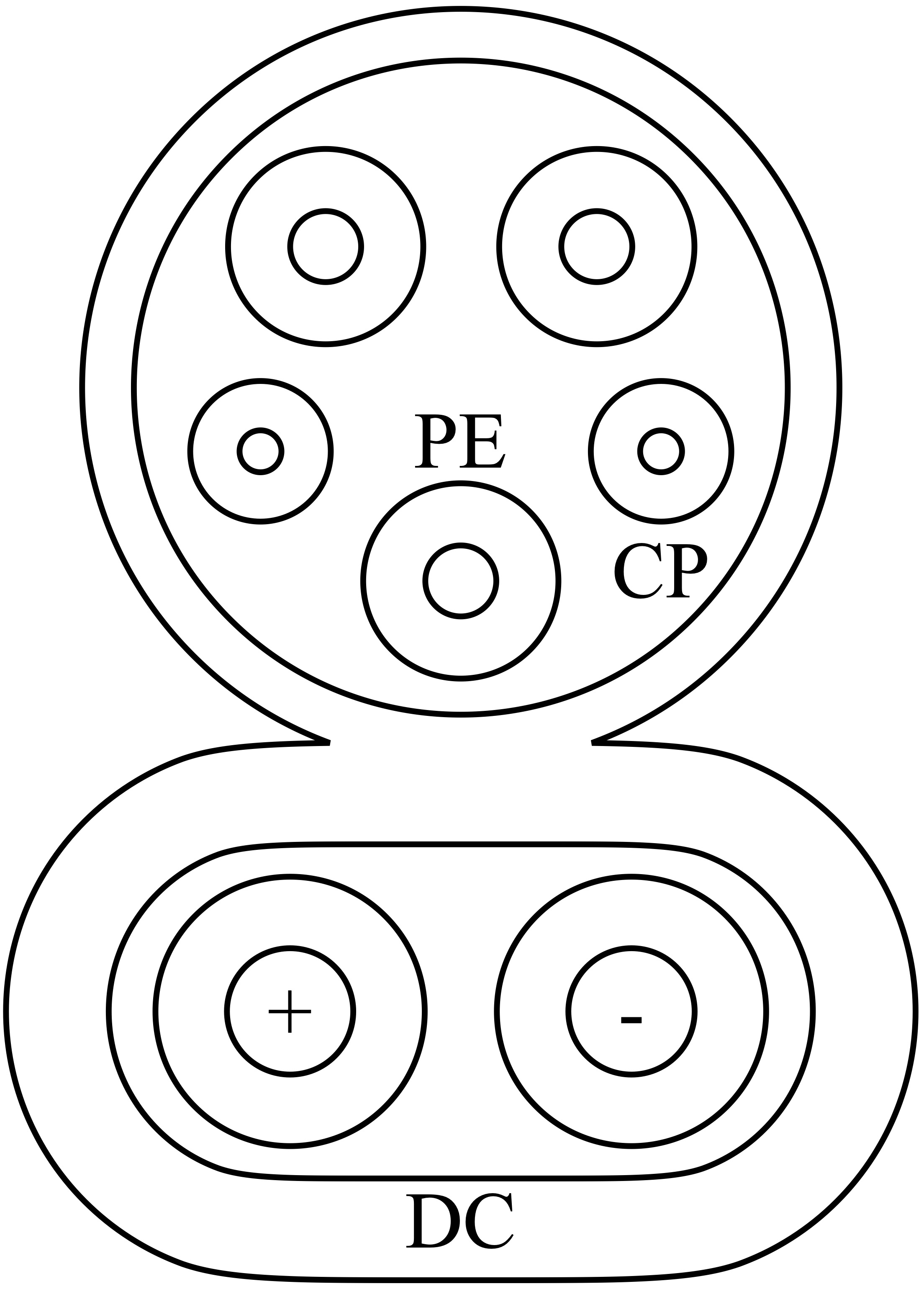}
		\caption{CCS Combo 1 (US).}
		\label{fig:ccs_combo1} 
	\end{subfigure}%
	\begin{subfigure}[b]{.499\linewidth}
		\centering
		\includegraphics[width=.6\textwidth]{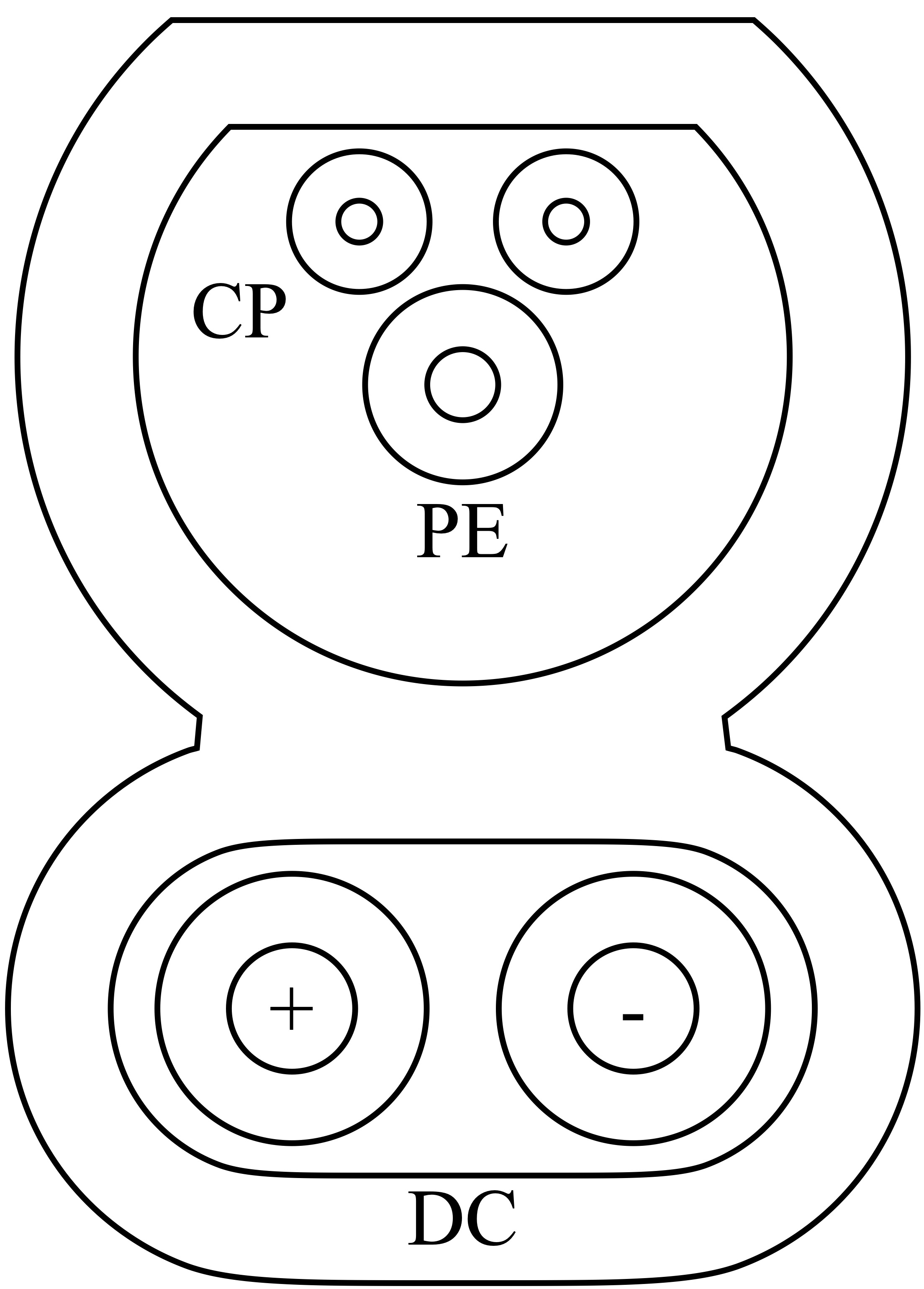}
		\caption{CCS Combo 2 (EU).}
		\label{fig:ccs_combo2} 
	\end{subfigure}%
	\caption{The two different plug layouts used by CCS in North America (Combo 1) and Europe (Combo 2) respectively.}
	\label{fig:ccs_plug_types}
\end{figure}

The charging technology standardized as the Combined Charging System  --- the name presented to a vehicle user --- is, in fact, a collection of multiple technical standards, assembled by the Charging Interface Initiative e.V. (CharIN e.V.), a non-profit association founded by a consortium of different stakeholders, such as automobile manufacturers, component suppliers and charging station vendors.
In this paper, we focus on communication between the EV and the EVSE, which is governed by the ISO 15118 series of standards\footnote{Previously by a simplified subset in DIN 70121, which provides limited capabilities but is functionally identical for underlying communication.}~\cite{iso15118-1}. 
Depending on the geographical region, CCS uses different plug types, which are illustrated in Figure~\ref{fig:ccs_plug_types}.
In North America Combo 1 and in Europe Combo 2 is used.
While the connector arrangement differs slightly between the two types, the underlying technology is the same.
As the name suggests, PLC usually uses the main power lines as a transmission medium.
However, in the case of CCS, the two separate wires, the Control Pilot (CP) and Protective Earth (PE) are used.
The actual direct current is delivered by the two large conductors at the bottom of the plug.

For backward compatibility, a basic communication scheme using a pulse-width modulation (PWM) protocol defined in the IEC 61851 standard~\cite{iec61851} is used when initializing the charging process. 
Upon successful initialization of this communication, a high-bandwidth IP link is established and control is handed to it. 
The vehicle and charger then engage in an application-layer protocol to control charging. 
This communication must be maintained throughout the charging session for a variety of reasons, the most important of which is fault detection. 
If the communication is lost, the ISO 15118 standard requires the CCS charging session to halt immediately and cease power transfer, upon which the EV and EVSE choose whether or not to attempt to start a new session from the beginning (or, in limited cases, continue charging using the basic pulse–width modulation protocol)~\cite{iso15118-2,iso15118-3}\footnote{This depends on various factors including the payment method used for the charging. We observed no instances of this fallback occurring in practice.}.

In addition to the main charging control loop, CCS also provides a range of additional capabilities using the IP link. 
One feature, known as `Plug \& Charge', enables automatic billing without the user authorizing a transaction manually~\cite{iso15118-1} --- and is currently appearing in the newest EVs and EVSE installations. 
Trials of `Vehicle-to-Grid' capabilities, in which vehicles can change their charging profile in response to power availability and even provide bidirectional power transfer, are at an advanced stage and these capabilities are expected to be publicly deployed shortly \cite{volkswagen_v2g}.

The Combined Charging System was initially intended only for light vehicles~\cite{iso15118-1}.
However, the lack of alternatives that provide additional benefits, such as higher charging capacity, led to the use in some much larger vehicles~\cite{minesmartferry, bydchargingsolutions}
Hence, derivative standards that are better suited to these uses have since been developed and are expected to be adopted in the future~\cite{charin_mcs}.
The SAE J3105 standard describes systems for charging electric buses using a range of connector options underpinned by ISO 15118~\cite{sae_j3105}.
Meanwhile, the new Megawatt Charging System (MCS) standard has been recently completed by the same standardization body as CCS, specifically designed to accommodate the needs of heavy vehicles, such as freight trucks~\cite{charin_mcs}. 
At the time of writing, trials have been run with 3.75 MW of charging power~\cite{charin_mcs}. 
We describe in Section~\ref{sec:impact} how our work is relevant for these standards, too.

\subsection{HomePlug Green PHY} \label{sec:hpgreenphy}

The high-bandwidth IP link used for communication in ISO 15118 is provided by the HomePlug Green PHY (HPGP) power-line communication technology~\cite{iso15118-3}. 
Power-line communication is a technology in which communication signals share cabling principally used for power transfer; with low-power high-frequency communication signals superposed alongside high-power direct- or alternating-current energy transmission~\cite{lampe2016}.
The HomePlug family of standards describe implementations of PLC intended for LAN communication, with HPGP as the derivation intended for embedded, industrial use-cases.
The HPGP standard is publicly available at no cost~\cite{alliance2013homeplug}, and is discussed in detail in~\cite{baker2019} and~\cite{bao2018threat}. 

In brief, HPGP uses an Orthogonal Frequency-Division Multiplexing (OFDM) physical-layer operating in the 2 -- 28 MHz band and modulating data onto (typically) 917 unmasked subcarriers spaced 24.414 kHz apart. 
The technology provides a set of communication modes that trade throughput vs. reliability with maximum speeds of 3.77, 4.92 (default), and 9.84~Mbps. 
The network is random access, with a Carrier-Sense Multiple Access scheme, with Collision Avoidance (CSMA/CA) to control medium access both within a network and when co-existing with other nearby HPGP networks. 
Although vehicle charging involves only two nodes in the HPGP network (EV and EVSE), the random-access nature of the technology still allows a situation in which both nodes transmit simultaneously. Furthermore, cross-talk from nearby charging stations could cause interference~\cite{baker2019}. 
To ensure that the communication is not affected by collisions, CSMA/CA is required.

The work in~\cite{baker2019} also describes a notable challenge with HPGP PLC: its propensity to leak communication signals from the charging cable. 
It has also been noted in~\cite{lampe2016} that the related HomePlug AV technologies are vulnerable to interference emitted in the same frequency range by broadcasting and wireless communication systems. 
This vulnerability was carried across to the HPGP technology and presents a risk of cross-talk between different vehicle-charger pairs in adjacent parking bays. 
It is possible that a vehicle would begin communicating, not with the charger that it is physically connected to, but with a nearby charger that happened to respond first, but is actually communicating via leaked signals. 
The HPGP specification implements the Signal Level Attenuation Characterisation (SLAC) protocol to mitigate this, in which the vehicle and charger exchange sounding messages to determine which are connected directly (thereby experiencing the least attenuation of their sounding messages) and which are being inadvertently overheard due to cross-talk~\cite{alliance2013homeplug}.

%% file: 04-ThreatModel/threatmodel.tex
\section{Threat Model}\label{sec:threat_model_goal}

\subsection{Goals}
The overarching goal of our considered attacker is the disruption of charging sessions for one or more EVs. 
We group possible intentions into three categories:

\ndssparagraph{Single Vehicle} 
In this case, a specific vehicle is targeted. 
This may be done as an attack on the owner, to make it difficult for them to travel, either at home or a remote location, thus exposing them to inconvenience or making them vulnerable to further physical attack.
Alternatively, it may be intended as an attack on the vehicle; immobilizing it at a remote location, from which it could be stolen if the driver leaves to obtain assistance.
Even in less sinister scenarios, disrupting the charge sessions of others would allow an attacker to take the newly-unlocked charge cable for use in charging their vehicle, or achieve faster charging by preventing other cars from sharing a load-balanced electricity supply.

\ndssparagraph{Fleet Denial} 
In this case, a specific organization is targeted and their vehicles immobilized en masse. 
This may be a delivery or transport business, in order to cause financial loss, harm supply chains or blackmail the operator for monetarization. 
Alternatively, the organization may represent a public service, such as buses, taxis, a police force, or ambulances, with the attack impacting local citizens.

\ndssparagraph{Widespread Disruption}
In this case, as many vehicles as possible are attacked, without regard to who owns or operates them. 
There may be an alternative attack rationale, such as harming the business of a charging service provider or influencing the local power grid through manipulation of the high power consumption of EVs and their proposed future use as bi-directional storage batteries. 
Given the high capacity of today's charging parks, which is expected to increase substantially in the next few years, the attack is an easy and effective way to control multiple megawatts of load.
It has been shown that maliciously controlled high-power devices can cause instabilities to the grid and are of great concern~\cite{soltan2018blackiot, sayed2022electric}.
Alternatively, the motivation may simply be spite --- blocking or vandalizing vehicle chargers due to a strong dislike of EV technologies has been documented in countries worldwide.

Overall, we consider an attacker who seeks to conduct their disruption with stealth, speed, and scalability. 
The simplest ways to interrupt charging are to operate an emergency cutout switch or to damage the vehicle, cable, or charger. 
However, such approaches require direct access, which is risky for perpetrators and mitigated with physical barriers or surveillance. 
Furthermore, the same process must be repeated for each targeted vehicle or charger, with commensurate repeated risks. 
In contrast, \textsc{Brokenwire} enables an adversary to disrupt charging sessions simultaneously at scale and from a safe, reasonable distance without interacting directly with the target(s). 
The attack can be performed from beyond the line of sight of potential CCTV cameras, concealing the attacker's presence.
In addition, physically-secured charging parks located behind a fence or wall are vulnerable.
The adversary can either execute the attack in person from a nearby location or deploy a device at the target site and control it remotely.

\subsection{Capabilities}

Since the entry barrier for carrying out the attack is low, our threat model considers a malicious actor with only access to off-the-shelf hardware that can easily be purchased online. 
At a minimum, this constitutes a software-defined radio and an antenna, with additional power amplifiers potentially being used to increase transmission power and, hence, range. 
We consider an attacker who can generate or capture the required attack signal --- which they may subsequently distribute to others to further reduce the barrier to entry.
Thus, even someone with little to no background in digital signal processing could retrieve an attack signal online and conduct the attack.

%% file: 05-Attack/attack.tex
\begin{figure}[t]
  \centering
  \includegraphics[width=0.99\linewidth]{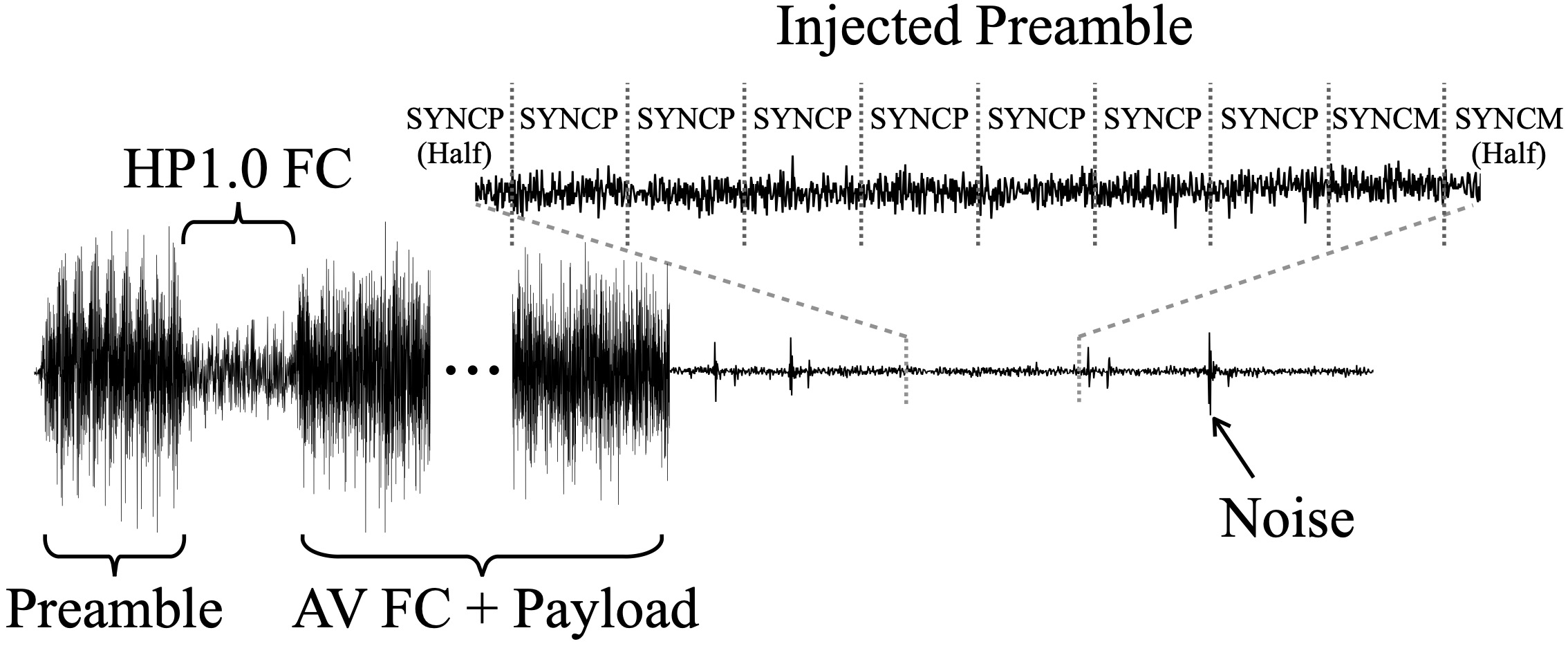}
  \centering
  \caption{Attack illustration. The injected preamble is not distinguishable from background noise. Only applying a correlation function on the captured data revealed the preamble position. Nevertheless, it caused a packet loss of around 80\%.} 
  \label{fig:attack_illustration} 
\end{figure}

\section{The \textsc{Brokenwire} Attack}

In this section, we give an overview of the technical details of the \textsc{Brokenwire} attack.
The attack involves repeatedly triggering the CSMA/CA mechanism that is required to be present in any standard-compliant implementation, such that neither vehicle (EV) nor charger (EVSE) ever have an opportunity to transmit. 
While this action alone could prevent communication indefinitely, it only needs to be applied for a few seconds in order to trigger a timeout in the higher layers of the communication protocol~\cite{iso15118-2,iso15118-3}. At this point the entire charging process is aborted and the attacker can stop broadcasting, making it only necessary to have temporary physical proximity to the victim.

\ndssparagraph{Exploiting CSMA/CA} \label{sec:exploiting-csmaca}
As mentioned in Section~\ref{sec:background}, the Combined Charging System uses HomePlug Green PHY PLC for communication during the charging session. 
The public HPGP standard defines Carrier-Sense Multiple Access with Collision Avoidance (CSMA/CA) as a channel access method. 
If a node wishes to transmit a message, it will check for other nodes already transmitting. 
In case an ongoing transmission is detected, the node will wait for a short, random period of time before attempting to transmit again~\cite{alliance2013homeplug}. 
This process is repeated indefinitely, until the transmission medium is idle, and the message can be transmitted.

The \textsc{Brokenwire} attack exploits this channel access mechanism to force the PLC modems at both nodes to back off and stop communicating. 
The attacker continuously transmits a recognizable signal, in this case a preamble, convincing any listening nodes that the channel is busy. 
The transmission is repeated indefinitely, such that both modems continue to wait and cannot transfer any data. 

Per the ISO 15118 standard, the communication session between the EV and EVSE is terminated if a short message timeout, the average in the standard being 1.5 seconds, is exceeded~\cite{iso15118-2,iso15118-3}. 
Once communication is lost, the charging session must cease immediately and a new session has to be initiated manually~\cite{iso15118-3}. 
This means that even after the attack ends, and the nodes can continue to communicate, the charging process will not automatically restart.

\ndssparagraph{Generating an Attack Signal}
In HPGP, the preamble is used to mark the beginning of a frame, to synchronize the receiver's clock with the transmitter and to permit channel state estimation. 
All frames begin with a standard HPGP preamble, which is sufficient to trigger the CSMA/CA mechanism in nodes that receive it. 
As such, \textsc{Brokenwire} uses a preamble waveform as an attack signal. 

The preamble is defined in the standard as a concatenation of repeated preamble symbols, each generated as follows:

$$ S[t] = \frac{10^{\frac{3}{20}}}{\sqrt{384}} \cdot \sum_{c \in C} cos \left( \frac{2 \cdot \pi \cdot c \cdot t}{384} + \psi(c) \right) $$

where $t$ is the preamble sample time step (for $ 0 \leq t \leq 384-1 $), $C$ is the set of unmasked subcarriers, $c$ is the subcarrier index and $\psi$ is a function mapping subcarriers to specific phase offsets defined in the standard~\cite{alliance2013homeplug}.

The concatenated symbols are then post-processed to fit one of two preamble formats, depending on compatibility mode. 
These modes differ only in the number of symbols used. 
In either case, the trailing symbols are inverted to assist with time synchronization and the entire preamble windowed to limit spectral leakage. 
The repeating structure, inversion of trailing symbols and windowing are visible in Figure~\ref{fig:attack_illustration}. 

As the preamble format is entirely documented, an adversary can generate a preamble from only public information. 
However, even this is not strictly required. 
Our empirical tests found that it was also highly effective to simply capture a small snippet of the communication between an EV and EVSE, extract the preamble and replay it. 
As such, the attack can be executed with little to no digital signal processing knowledge.

\ndssparagraph{Injecting the Attack Signal}
The attack signal can be trivially injected with physical access.
However, as shown by~\cite{baker2019}, the charging cable acts as an unintentional antenna that leads to electromagnetic emanation.
At the same time, this phenomenon makes the charging cable susceptible to electromagnetic interference.
Since the cable is unshielded, electromagnetic waves can easily couple onto the wires within it. 
While the PLC uses differential signaling over two wires, any asymmetries in the two pathways lead to some signal still being retained. 
By transmitting the attack signal over-the-air, an attacker can therefore cause sufficient coupling on the charging cable of a victim EVSE for it to correctly detect the injected preambles. 

This makes the use of a CSMA/CA attack particularly powerful. 
While devices are actively designed to resist noise, they are designed to detect faint preambles efficiently.
We found that broadcasting a preamble over-the-air, in the same frequency band in which PLC is operating, couples well onto the cable and is picked up by the PLC modems.
In accordance with the standard, if the signal-to-noise (SNR) ratio of the preamble is $\geq 2$~dB, the preamble is correctly recognized and interpreted as the start of a packet which indicates an ongoing communication~\cite{alliance2013homeplug}.

%% file: 06-Evaluation/evaluation.tex
\section{Lab Evaluation} \label{sec:attack_evaluation}

In this section, we present the results of our experiments with the attack in a controlled laboratory environment. Later, in Section~\ref{sec:real_world_evaluation}, we describe real-world testing.

\subsection{Experimental Setup} \label{sec:experimental_setup}

\begin{figure}[t]
  \centering
  \includegraphics[width=0.8\linewidth]{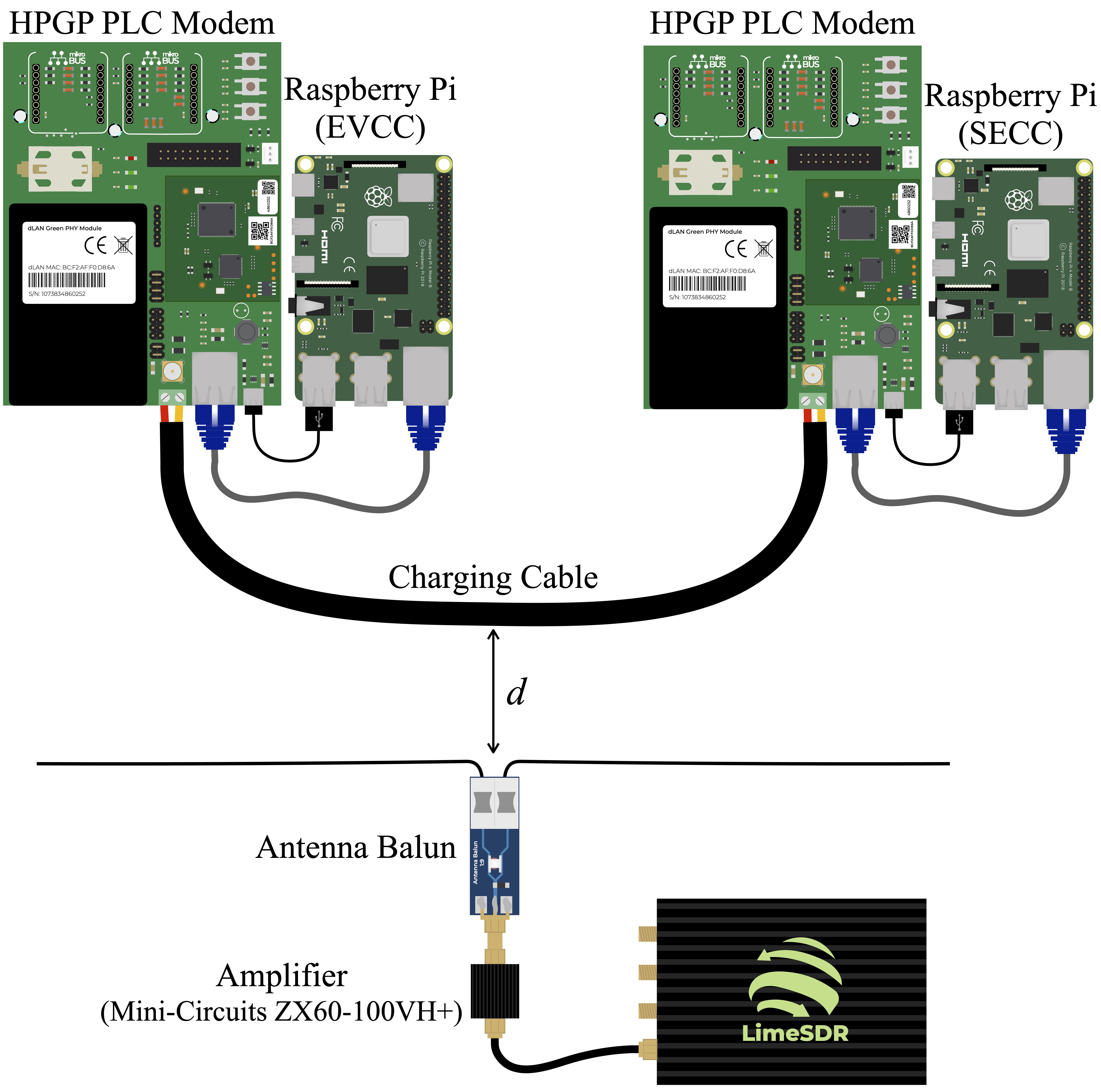}
  \centering
  \caption{An overview of the experimental setup used to evaluate the attack under controlled lab conditions.} 
  \label{fig:experimental_setup} 
  \vspace{-0.7cm}
\end{figure}

We evaluated the attack under controlled laboratory conditions, using a testbed composed of two PLC HomePlug Green PHY evaluation boards from Devolo (dLAN Green PHY eval board II)~\cite{devolo_devboards}.
The boards were equipped with the same Qualcomm QCA7000 chips, as they are used for communication in EVs and DC high-power charging stations~\cite{qualcom_qca7k}.
The two HPGP evaluation boards were connected via a 4~m long copper wire with two flex cores (2 $\times$ 1.5~mm\textsuperscript{2}), one for the Control Pilot and one for Protective Earth, the same cable structure found in CCS charging cables.
The cable length corresponded roughly to the same length found at most DC rapid chargers.
Each PLC modem was connected to a RaspberryPi via Ethernet, which mimicked the Electric Vehicle Communication Controller (EVCC) and Supply Equipment Communication Controller (SECC).
The communication controllers are responsible for the higher-level communication (HLC). In our case, they ran IPerf, as described in the subsequent section.
As such, any device that can handle the HLC would be suitable for this task.

On the attacker side, we used a LimeSDR as the transmitter connected to a 1~W amplifier (Mini-Circuits ZX60-100VH+).
We used GNURadio to operate the LimeSDR and emit the malicious attack signal.
Our antenna was a low-cost solution composed of a balun (1:9) with a dipole made from two simple 24AWG copper wires. 
Our antenna was electrically-short (7~m) for the frequency band, so we expect the attack to be even more effective with a dipole optimized for maximum gain (\( \frac{5}{4} \lambda \), where \( \lambda \) is the wavelength)~\cite{kraus2002antennas}.
The entire experimental setup is depicted in Figure~\ref{fig:experimental_setup}.

\subsection{Method}\label{sec:method}

\begin{figure*}[t]
	\centering
	\begin{subfigure}[b]{.499\linewidth}
		\centering
		\includegraphics[width=.9\textwidth]{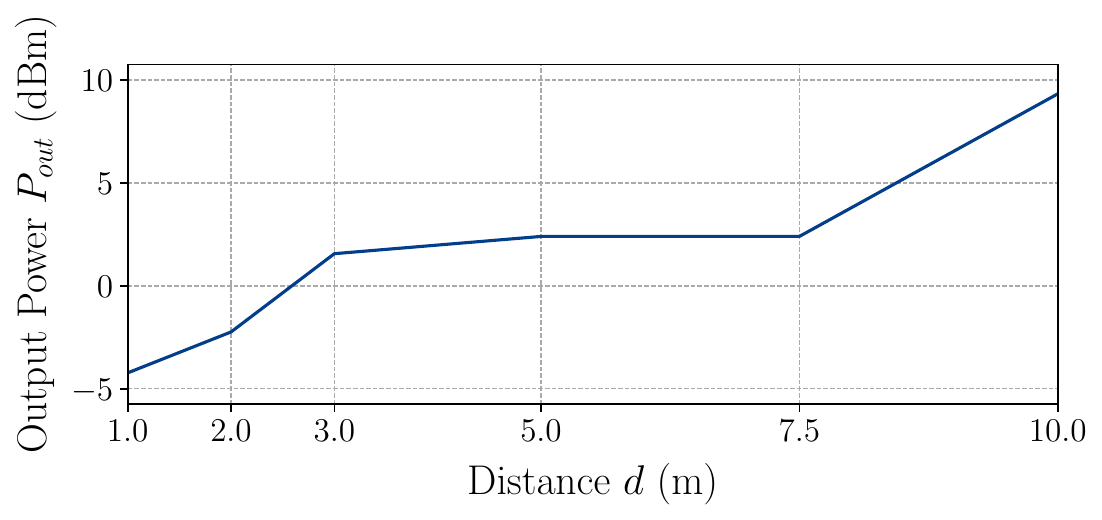}
		\caption{Results in dBm.}
		\label{fig:req_power_dbm} 
	\end{subfigure}%
	\begin{subfigure}[b]{.499\linewidth}
		\centering
		\includegraphics[width=.925\textwidth]{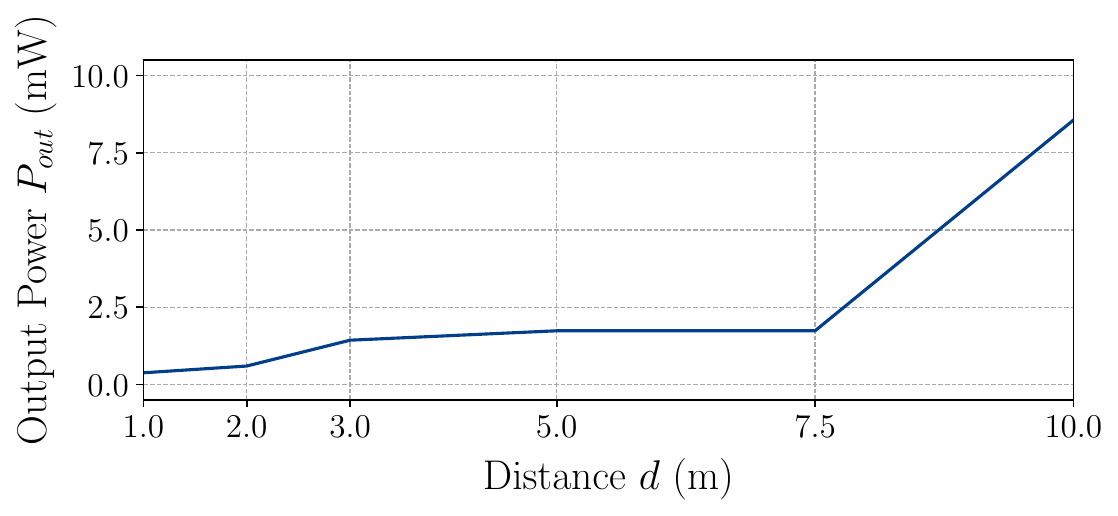}
		\caption{Results in mW.}
		\label{fig:req_power_mw} 
	\end{subfigure}%
	\caption{Results of the distance experiments in lab conditions. Minimum power required to cause a packet loss of 100\%.}
	\label{fig:disruption_results_lab}
\end{figure*}

We evaluated the effectiveness of the attack on network throughput using IPerf, an open-source software tool for performance testing.
We set up a UDP communication between the two RaspberryPis, whereby the IPerf server was running on the EVCC and the IPerf client was executed on the SECC.
Under normal operation, we observed no packet losses and a maximum transmission rate of 833 packets per second for a throughput of 5~Mbps.
We considered the communication link as offline and the attack as successful, once IPerf reported an unsuccessful establishment of a connection or a packet loss of 100\%.

In order to determine the minimum required transmission power for a successful attack for a given distance $d$, we iterated over the gain settings of the LimeSDR (14 -- 64) with a step size of 2.
For each gain setting, we broadcast a preamble for 20 seconds and measured the packet loss during this period.
The center frequency of the LimeSDR was set to 17 MHz and the sample rate to 25~MSPS.
While the sample rate of 25~MSPS was not enough to cover the entire spectrum used by HPGP, it was sufficient for our attack.

To emphasize the simplicity and effectiveness of our approach, we replayed a captured preamble rather than using one generated from scratch. 
We argue that this yields conservative results as the captured preamble will have experienced additional distortion and should therefore be less likely to be recognized by the PLC modems. 
Moreover, it highlights the fact that the adversary does not have to know how to generate the preamble making the attack even easier to execute.

We want to note that although the experiments were as controlled as possible, some parameters were out of our control.
For example, the noise that coupled onto the PLC lines from other devices in the building varied widely.
To compensate for these variations, we conducted multiple runs and calculated the mean power required for the attack to work.
Finally, we repeated the experiment for different distances $d$ between the charging cable and the antenna.
To ensure reproducibility, we automated our testing by using a Python script that iterates over the different output powers\footnote{The evaluation source code is available at: \url{https://github.com/ssloxford/brokenwire}.}.

\ndssparagraph{Measuring Output Power}
The output power of the LimeSDR was not calibrated and the gain was only given as a unitless number.
In order to report accurate values for the required transmission power, we measured the actual output power of the LimeSDR for each gain setting.
We directly connected the LimeSDR to an RF power meter and transmitted the preamble used in our evaluation. 
\subsection{Results}

The efficiency of the \textsc{Brokenwire} attack becomes apparent when examining the results presented in Figure~\ref{fig:disruption_results_lab}.
The graphs show the minimum required transmission power for various distances to cause a packet loss of 100\% or IPerf to throw a ``No route to host.'' exception.
As previously mentioned, the signal-to-noise ratio of the injected preamble only needs to be $\geq 2$~dB to be correctly interpreted as the start of a packet.
Unanimously with this requirement, it is less surprising to find that even an extremely low output power is sufficient to induce a strong enough preamble into the wire to interfere and disrupt communications.
Although the required transmission power increased substantially with greater distance, 10~mW was still sufficient to disrupt the communication of the testbed from 10~m away.

\subsection{Effectiveness across Multiple Stories}

The advantage of using electromagnetic emanation as an attack vector is that no physical access as well as line-of-sight to the target is necessary. 
To demonstrate the capabilities of our attack, we conducted it in a limestone building with thick walls and floors, and a ceiling height of about 3.5~m. 
We positioned the attacker on the ground floor of the building and placed the charging testbed one story above. 
The components were not directly aligned above each other, but slightly offset, resulting in a distance of approximately 6~m between the antenna and the PLC modems, including a ceiling approximately 20~cm thick.
In line with the other lab experiments, we ran a UDP IPerf session between the RaspberryPis that were connected via the PLC modems.
We then increased the gain of the LimeSDR with a step size of two until IPerf reported a packet loss of 100\% or the unsuccessful establishment of a connection.
We repeated the experiment multiple times and averaged the required output power.
We found that an output power of around 100~mW was sufficient to disrupt the communication.
While precise ranges and power budgets will vary between environments, we believe that this result amply demonstrates that the attack can be conducted beyond a physical barrier and, given the nature of the test building, that it approximates even a multi-story parking lot. 

\subsection{Comparison to Broadband Noise Jamming}\label{sec:evaluation_of_noise}

HomePlug Green PHY was designed to withstand noisy environments~\cite{alliance2013homeplug}.
Using the ROBO mode, which transmits the data redundantly on multiple subcarriers, and the usage of QPSK as a modulation scheme, make it robust even under severe communication channel conditions.
To emphasize the efficiency of our attack and demonstrate the robustness of HPGP against noise, we compared the required power of the preamble injection to the signal required for successful broadband noise jamming.
As the name suggests, noise is emitted simultaneously in the entire spectrum in which PLC is operating.
We used the same experimental setup and method as described in Section~\ref{sec:experimental_setup} with a fixed distance between the target and the attacker of 1~m.
However, instead of broadcasting a preamble, we just emitted Gaussian White Noise.
With the maximum output power of the Mini-Circuits ZX60-100VH+ (1~W), we could not observe any degradation of the connection quality whatsoever.
Hence, we replaced the 1~W amplifier with a Kalmus 161C amplifier (max. 100~W) and repeated the experiment for transmission powers up to 20~W. 
We still observed no effect on the connection.

The results are in line with our expectations and validate our hypothesis in Section~\ref{sec:exploiting-csmaca} that a preamble injection attack substantially outperforms noise as a disruption technique. 
Not even a signal with three orders of magnitude more power was enough to disrupt the communication. 

\subsection{Predicting Attack Range}

\begin{figure}[t]
	\centering
	\begin{subfigure}[b]{0.99\linewidth}
		\centering
		\includegraphics[width=\textwidth]{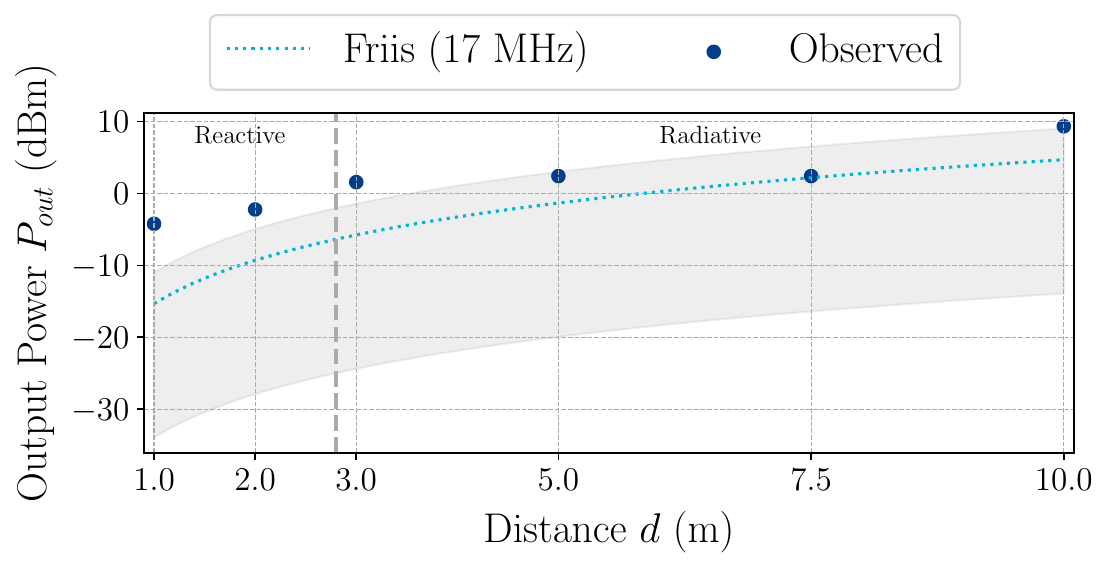}
	\end{subfigure}%
	\caption{Comparison of observed results in our testbed against predictions from the Friis equation, for signal $\geq 2$~dB greater than noise of 2.45~mV. Shaded region represents the range of predictions across the 2 -- 28~MHz frequency range. Dashed vertical line represents transition between near-field regions. The far-field region is beyond the range of the plot.}
	\label{fig:disruption_results_vs_maths}
\end{figure}

A range of mathematical methods exist for estimating received power over a wireless link. 
However, the majority of methods employ simplifying assumptions that may not always hold valid. 
The Friis path-loss equation is often used to estimate expected power levels in wireless security contexts~\cite{friis1946note}. 
However, the equation is only well-defined for signals in the far-field of transmission and is frequency dependent. 
At the comparatively low frequencies used by HPGP PLC signals, the far-field begins at least 10.7~m away from the transmitter and the difference in predicted power at a receiver changes substantially when considering the lowest or highest parts of the signal bandwidth. 

Given these concerns, we investigated whether our observed results bore any similarity to those predicted by the Friis equation. 
To do so, we measured the ambient noise level on the charging cable in our testbed, finding it to be 2.45~mV. 
From this, we computed the minimum preamble detection threshold stated in the HPGP standard ($\geq2$ dB compared to noise). 
We then applied a rearranged Friis equation to determine the transmission power required to achieve this received power level. 
Full details of the calculations are provided in Appendix~\ref{app:friis-comparison-details}.

Figure~\ref{fig:disruption_results_vs_maths} shows a comparison of the results observed in our testbed with predictions made using the Friis equation. 
We noted that, while the Friis estimates provide a wide range of powers, depending on the selected frequency, the estimates made using the upper end of the band were close to our observed results. 
This is consistent with the observations in~\cite{baker2019}, showing that the higher part of the band appeared most prone to radiating from the charging cable (and thus, we expect most prone to ingress as well). 
The predicted values were furthest from the observed ones in the reactive near-field region, but became closer in the radiative near-field. 
We tentatively suggest that the predictions would continue to be accurate beyond our furthest measured distance.

\section{Real-world Testing}\label{sec:real_world_evaluation}

\begin{figure*}[t]
	\centering
	\begin{subfigure}[b]{.195\linewidth}
		\centering
		\includegraphics[width=.92\textwidth]{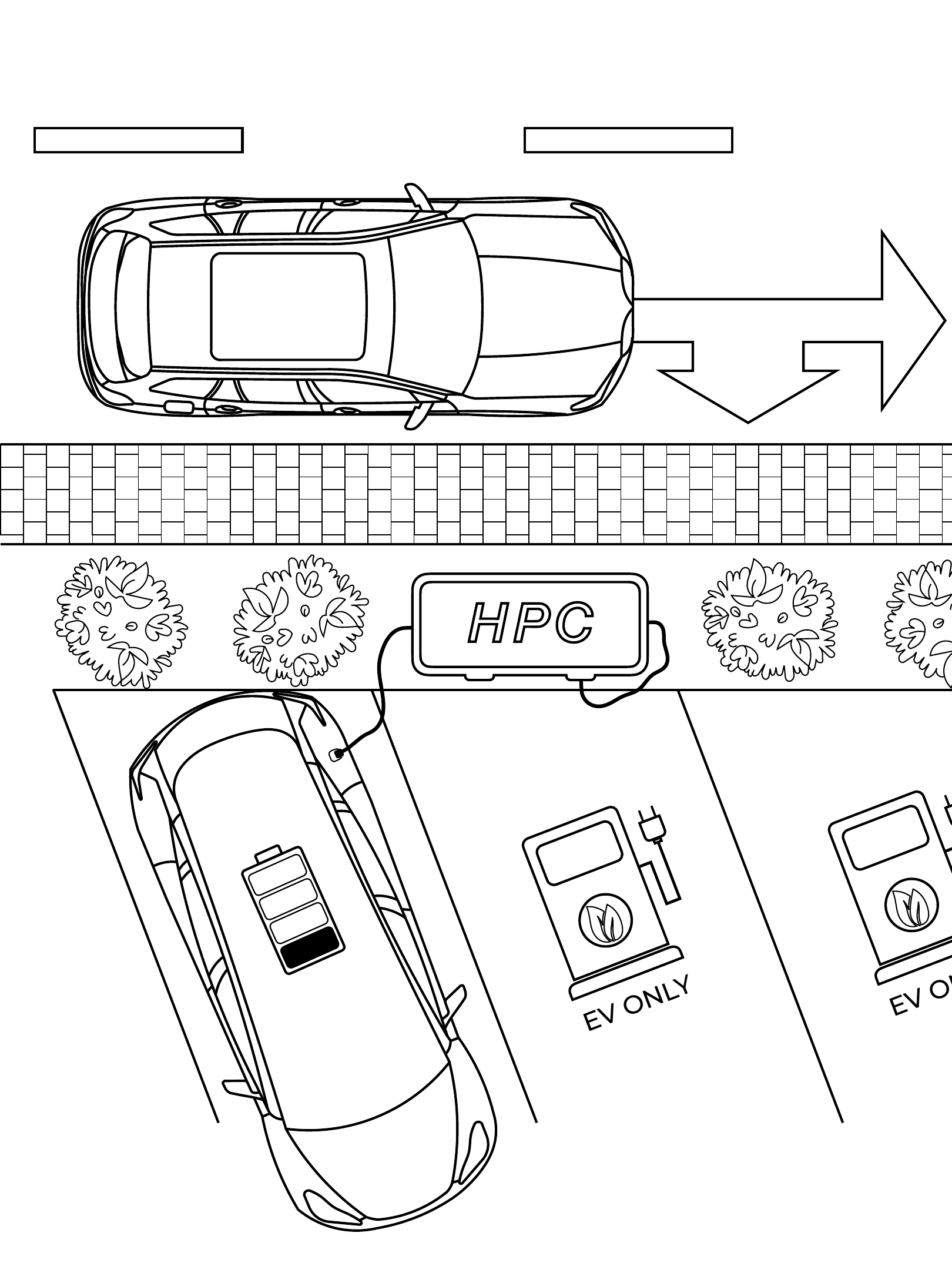}
		\caption{Scenario 1.}
		\label{fig:scenario_1} 
	\end{subfigure}%
	\begin{subfigure}[b]{.195\linewidth}
		\centering
		\includegraphics[width=.92\textwidth]{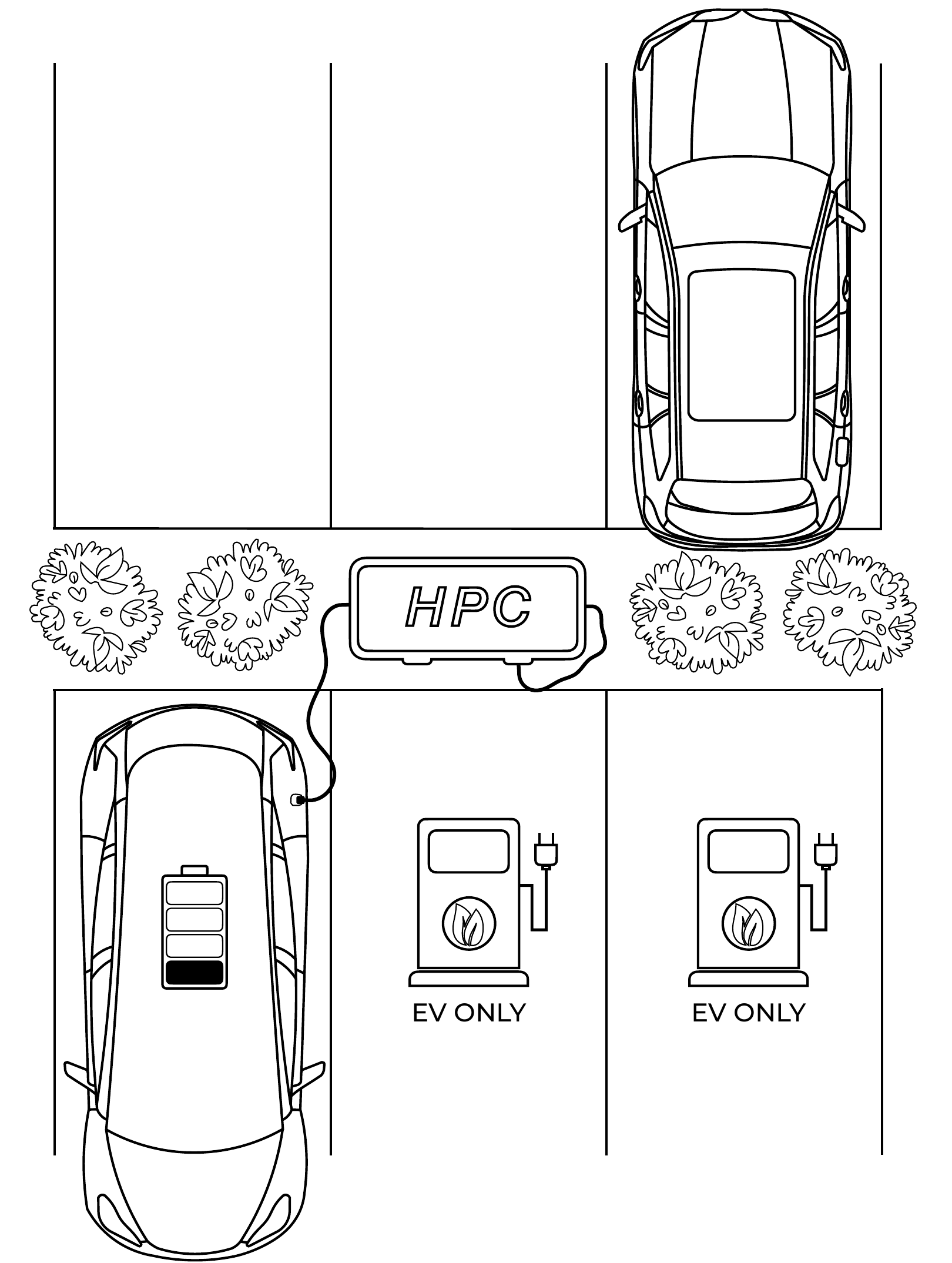}
		\caption{Scenario 2.}
		\label{fig:scenario_2} 
	\end{subfigure}%
	\begin{subfigure}[b]{.195\linewidth}
		\centering
		\includegraphics[width=.92\textwidth]{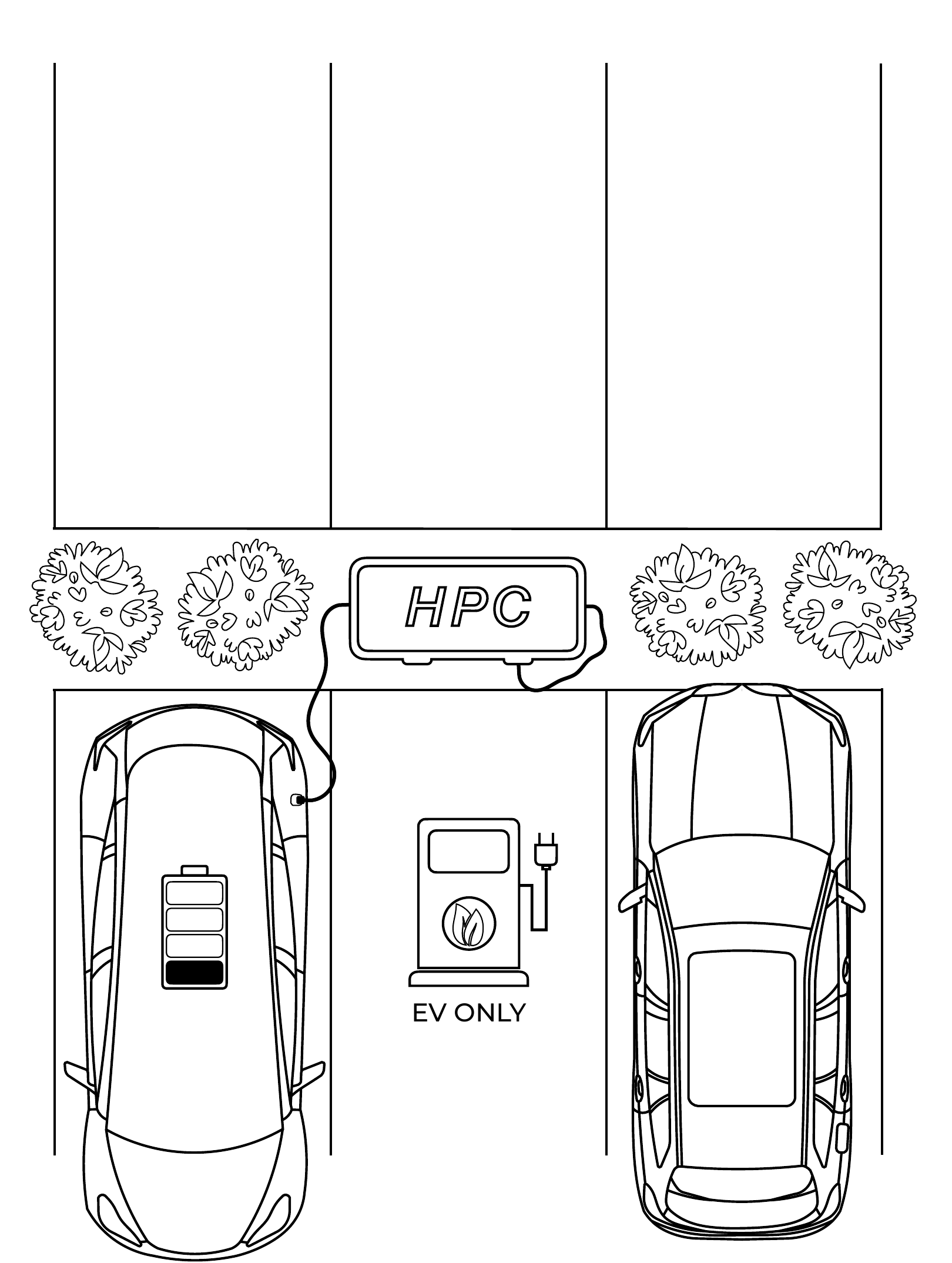}
		\caption{Scenario 3.}
		\label{fig:scenario_3} 
	\end{subfigure}%
	\begin{subfigure}[b]{.195\linewidth}
		\centering
		\includegraphics[width=.92\textwidth]{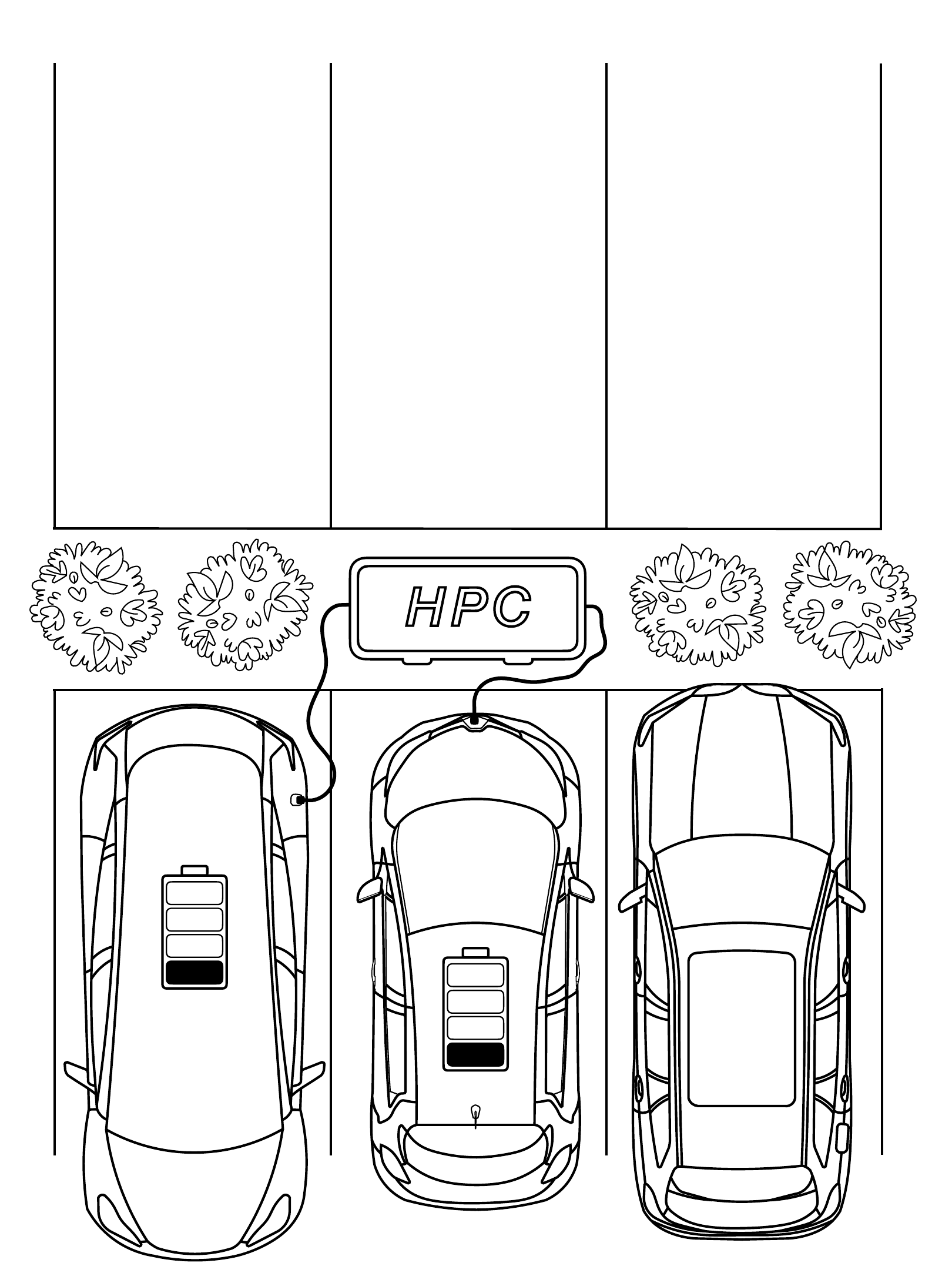}
		\caption{Scenario 4.}
		\label{fig:scenario_4} 
	\end{subfigure}%
	\begin{subfigure}[b]{.195\linewidth}
		\centering
		\includegraphics[width=.92\textwidth]{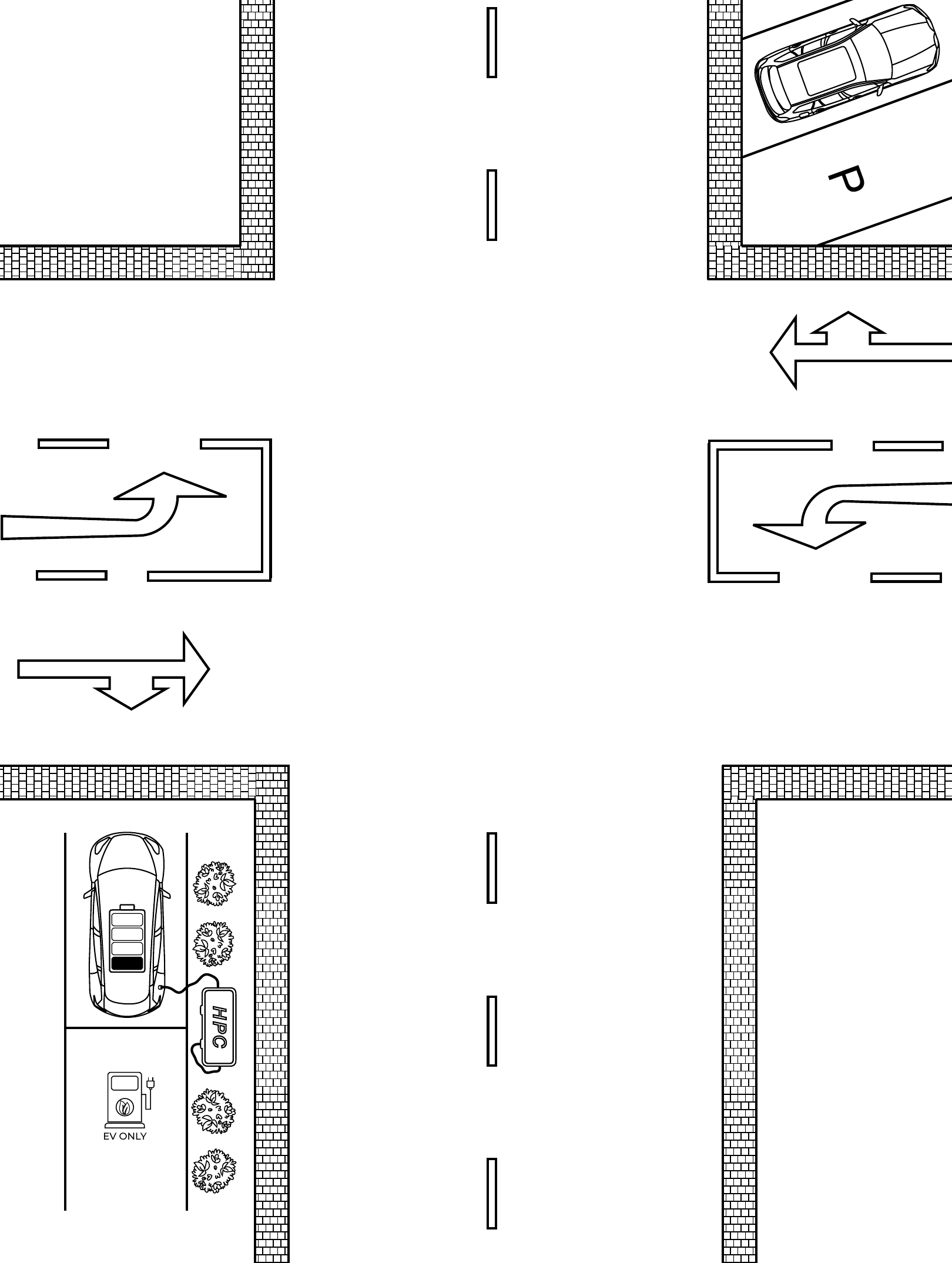}
		\caption{Scenario 5.}
		\label{fig:scenario_5} 
	\end{subfigure}%
	\caption{Example scenarios that we tested during our real-world evaluation. In all of the scenarios, the victim is represented by the car(s) connected to the charging station and with the battery symbol. The vehicle that is not charging is the adversary.}
	\label{fig:real_world_scenarios}
\end{figure*}

Following our experiments in a controlled lab environment, we subsequently examined the effects of the \textsc{Brokenwire} attack on real charging sessions.
We tested the attack on a broad range of passenger cars from different manufacturers, spanning across different classes, price ranges, and charging capacities.
All eight cars were equipped with a CCS charging port, either Combo 1 (US) or Combo 2 (EU), and followed the ISO 15118 or the DIN 70121 standard. 
Table~\ref{tab:tested_vehicles} gives an overview of the cars we tested.
In addition, we evaluated a total of 20 DC charging stations from various providers and manufacturers. 
We did not exhaustively test every combination of vehicles and charging stations, but tested multiple vehicles with most chargers. 
In one case, we successfully tested with three vehicles charging simultaneously on identical charging stations.

\begin{figure}[t]
  \centering
  \includegraphics[width=0.99\linewidth, trim = 12mm 17mm 0 33mm, clip]{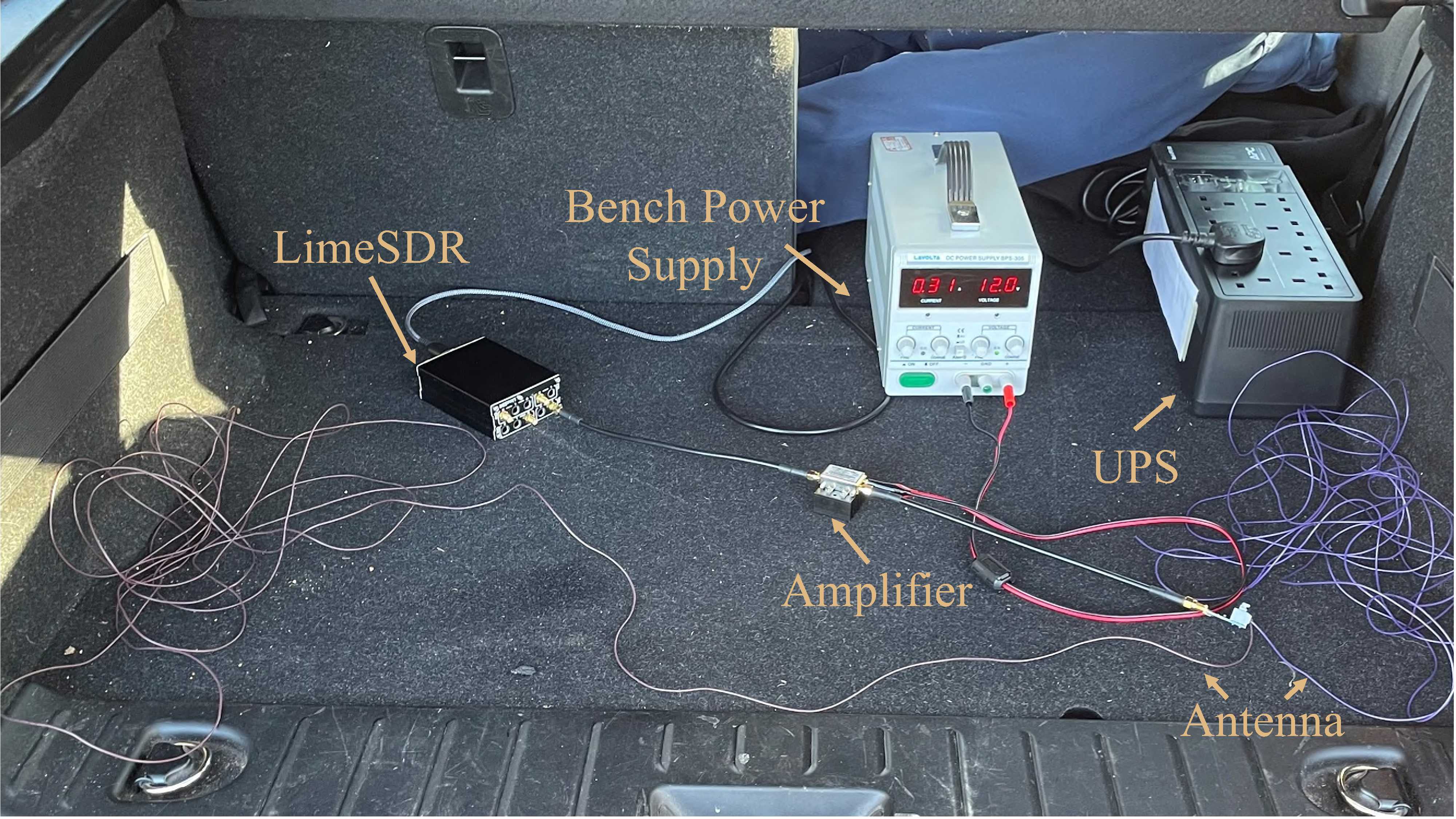}
  \centering
  \caption{The equipment we used during the real-world experiments. The hardware, including the antennas, was installed in the back of the trunk and powered via a UPS.} 
  \label{fig:real_world_setup} 
\end{figure}

\begin{table}
	\centering
	\footnotesize
	\caption{Overview of the eight tested vehicles.}
	\begin{tabular}{llll}
		\toprule
		\textbf{Vehicle} & \textbf{Class} & \textbf{Price (\$)} & \textbf{Charging Capacity}\\\hline
		Vehicle A & Subcompact & 50,000 & 50 kW\\
		Vehicle B & Compact SUV & 85,000 & 150 kW\\
		Vehicle C & Shooting Brake & 150,000 & 270 kW\\
		Vehicle D & Subcompact & 20,000 & 50 kW\\
		Vehicle E & Mid-size Sedan & 50,000 & 120 kW\\
		Vehicle F & Mid-size SUV & 70,000 & 150 kW\\
		Vehicle G & Compact & 45,000 & 125 kW\\
		Vehicle H & Compact & 32,000 & 50 kW\\\bottomrule
	\end{tabular}
	\label{tab:tested_vehicles}
\end{table}

\subsection{Method}

We evaluated the \textsc{Brokenwire} attack in various scenarios that replicate how it could be carried out in a real-world attack.
Figure~\ref{fig:real_world_scenarios} illustrates five of the different scenarios that we examined. 

\ndssparagraph{Scenario 1}
In this scenario, we simulated a drive-by attack, in which the adversary aims to disrupt the charging session of vehicles that are parked close to a public road.
To achieve the maximum possible coverage, we mounted the antenna onto the roof of the attacking car on the site closest to the target vehicle.
We then drove slowly (approx. 10~mph/15~kph) past the charging vehicle.
A drive-by attack could be used by an adversary to fulfil the goal of fleet denial or widespread disruption as described in Section~\ref{sec:threat_model_goal}.
Since only temporary physical proximity is required, so-called wardriving would be possible.
This makes the attack stealthy and challenging to detect.

\ndssparagraph{Scenario 2} 
This scenario mimicked a situation that can often be found on public car parks, for example, at supermarkets. 
We parked the attacking car, with the antenna coiled in the trunk, in a parking bay on the opposite side of the victim.

\ndssparagraph{Scenario 3}
Similar to Scenario 2, we emulated a public car park. 
However, this time, we parked the attacking vehicle parallel to the victim.
In line with the previous scenario, we kept the antenna coiled in the trunk.

\ndssparagraph{Scenario 4}
Scenario 4 had the same settings as Scenario 3.
However, we parked another vehicle between the attacker and the victim.
This enabled us to evaluate the attack for a setting where there was no direct line-of-sight between the attacker and the victim.
At the same time, it allowed us to test the attack on multiple cars at once. 
This scenario, along with the previous two, particularly reflects an attacker seeking to access an in-use charging cable or avoid sharing the available current with others.

\ndssparagraph{Scenario 5}
In Scenario 5, we simulated an attack from a distance to hide the physical presence of the adversary.
The victim was charging close to a large intersection and the adversary was located on the opposite side of the intersection.
This setting can, for example, be used to target an EV fleet at a commercial site/depot with CCTV surveillance.

The real-world evaluation was performed with the same equipment used for the laboratory experiments.
However, the amplifier was powered from an Uninterruptible Power Supply (UPS) and a bench power source.
The setup used is depicted in Figure~\ref{fig:real_world_setup}.
This research prototype cost less than \$1,000 and is easily contained in the vehicle. 
However, we later achieved a smaller, lighter setup by exchanging the UPS and bench power source with a small USB power bank and a step-up converter.
The capacity of the battery (10~Ah) was sufficient to run the attacking hardware for more than 10 hours and could easily be carried in a rucksack or mounted on a drone.
We are confident that further optimizations could be applied for size and cost that would allow an adversary to plant the device in a remote location and continuously disrupt an entire area.
Depending on the tested scenario, the antenna was either placed within the attacking vehicle, mounted on the bodywork, or positioned on the ground immediately next to it. 

\subsection{Observations}

The \textsc{Brokenwire} attack was successful in disrupting charging in every observed case --- for any combination of EV and EVSE we tested, and independently of the CCS plug used (Combo 1 and Combo 2).
Substantial variations were observed in effective range and power requirements in each case. 
However, once a suitable power level had been selected for the environment, charging sessions were terminated immediately.

\ndssparagraph{Required Output Power}
\begin{figure}[t]
  \centering
  \includegraphics[width=0.98\linewidth, trim = 0 105mm 0 115mm, clip]{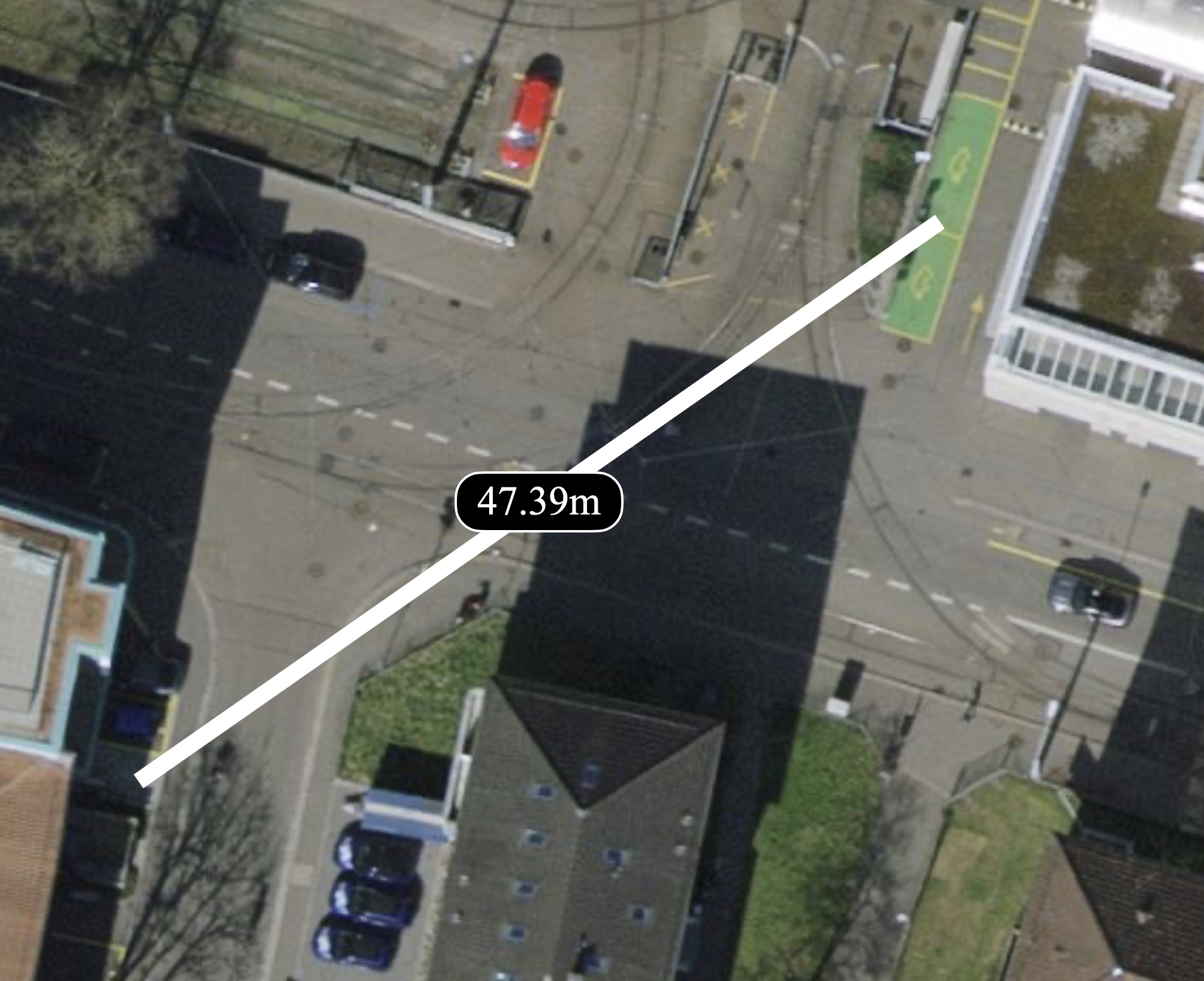}
  \centering
  \caption{Distance measurement for Scenario 5 via zoom.earth. The target charges at the green parking bay, while the attacker is located on the other side of the intersection.} 
  \label{fig:distance_attack} 
\end{figure}
In contrast to the lab experiments, the conditions during the real-world experiments were largely uncontrolled, which led to less predictable results.
In particular, we noticed that in some cases, a much higher transmission power was required to successfully terminate the charging session.
For example, while we successfully disrupted the testbed from 10~m away using less than 10~mW of transmission power (see Figure~\ref{fig:req_power_mw}), disrupting Vehicles B and C from a distance of 10~m in a setting similar to Scenario 2 required around 600~mW.
In contrast, at another charging site, we stopped vehicle D from charging using only the LimeSDR \textit{without the amplifier} from 1.5~m away, which according to our measurements, corresponded to 0.3~mW of output power.
At another EVSE, we replicated Scenario 5 and successfully disrupted an ongoing charging session of Vehicle D from 47~m away with just under 700~mW.
The exact arrangement of this scenario is depicted in Figure~\ref{fig:distance_attack}.
We believe that the large variations in the power requirements are due to several parameters that cannot be controlled by the adversary, such as electromagnetic interference from other devices, differences in signal propagation and noise generated by the charger or the vehicle itself.
Nevertheless, the results from our real-world evaluation are largely in agreement with the results and observations from our laboratory experiments, showing that the attack is effective and easy to execute.

\ndssparagraph{Disrupting Multiple Vehicles}
Since large charging hubs, where multiple vehicles can be charged simultaneously, are becoming more and more common, we also evaluated the effects of the \textsc{Brokenwire} attack in such a setting.
The arrangement of our test was similar to Scenario 4.
However, we tried to interrupt the charging sessions of three cars at once.
We connected the vehicles to two separate charging stations and successfully disrupted all charging sessions at about the same time.

To further demonstrate the scalability of the attack and examine how the signal propagates when the line-of-sight is occluded by other vehicles, we conducted the attack again in a large car park with three rows of cars between the attacker and the victim. 
While the vehicles that occluded the line-of-sight were parked on normal parking bays without chargers and only the victim cars were charged and disrupted, the layout of the car park mimicked a real, multi-unit charging hub, illustrating how the attack would affect other vehicles in the vicinity.
In this specific scenario, a total of 20 vehicles were parked in the range of the attack signal.
Therefore, we are positive that any EV within that range would have been affected by the attack.
The results also prove that the attack works against occluded vehicles.

\ndssparagraph{Antenna Alignment}
We noticed that the alignment of the antenna is crucial for the effectiveness of the attack.
The attack worked reliably and within moderate distance when the antenna was coiled in the closed trunk, as shown in Figure~\ref{fig:real_world_setup}.
However, for the same output power, the maximum possible attack range increased significantly when the antenna was stretched out.
Moreover, we observed that the signal is attenuated when the trunk is closed.
Moving the antenna outside increased the attack range even further. 
Nevertheless, our findings indicate that even a coiled antenna is adequate to conduct a successful attack.
This means the antenna can easily be hidden in, for example, a backpack.

\ndssparagraph{Error State}
\begin{figure}[t]
  \centering
  \includegraphics[width=0.98\linewidth, trim = 0 270pt 0 110mm, clip]{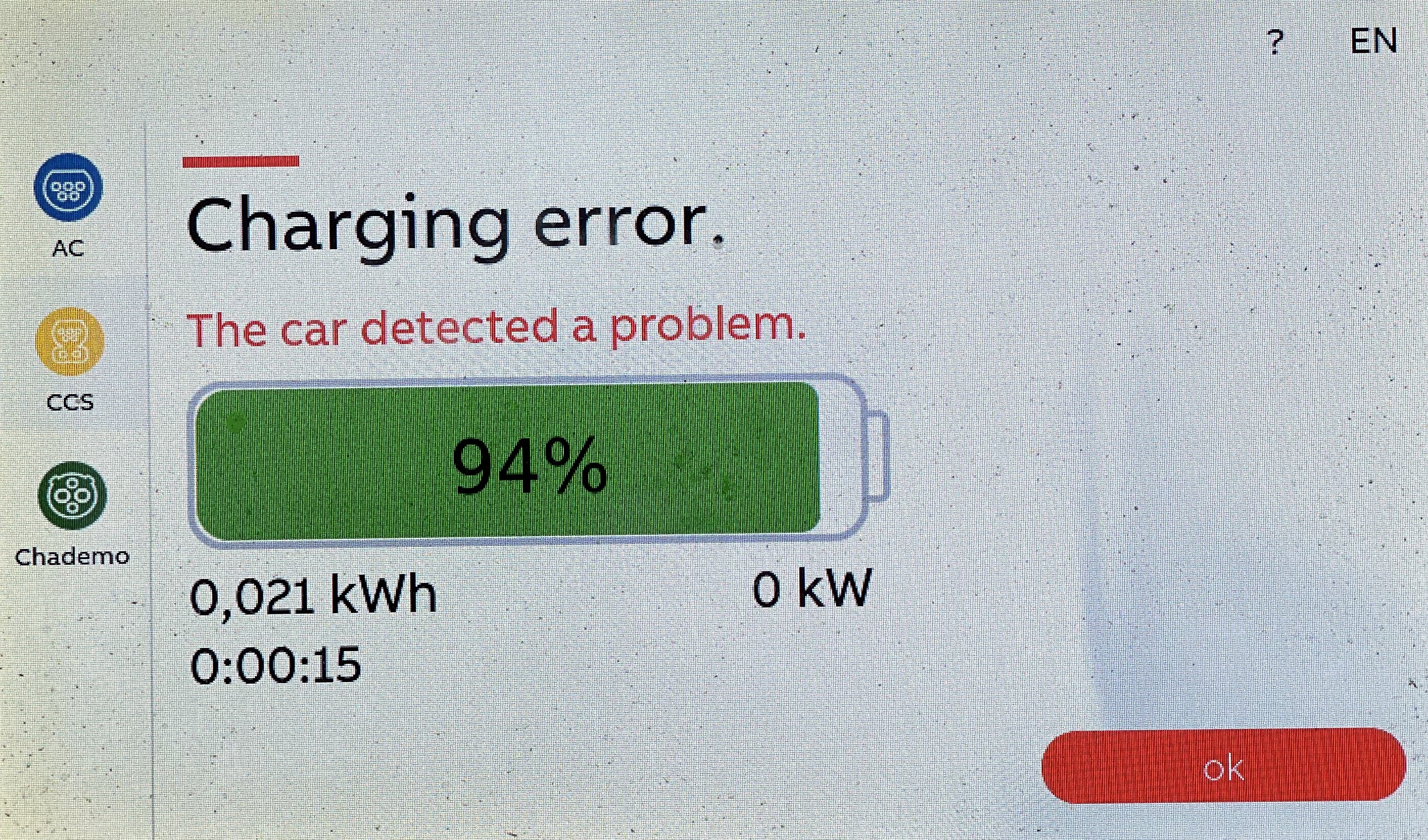}
  \centering
  \caption{The error message displayed on a tested charging station shortly after starting the broadcast of the preamble.} 
  \label{fig:evse_error} 
\end{figure}
In all of our tests and independent of the vehicle and charging station, we observed that after the successful disruption of the charging communication, the car and the charger switched into an error state.
An example error message displayed on one of the tested chargers is depicted in Figure~\ref{fig:evse_error}.
Even though the ISO 15118 standard provides the option of continuing the charging using the basic PWM communication~\cite{iso15118-3}, it appears that none of the tested vehicles implemented this feature.
To continue charging, it was necessary to manually start an entirely new charging session from scratch. 
This means we had to unplug the charging cable from the vehicle and plug it back into the charging station, wait for a short period, plug the cable into the car again, authenticate the charging via the preferred authentication method and wait for the charging to start.
While we have not observed the behavior for Plug \& Charge, we are confident that the car will not automatically continue charging once the communication link is re-established.
We consider this to be a safety feature since the communication is crucial for operating the equipment safely.
To our surprise, one of the tested charging stations did not show an error at all.
Instead, the charger stated that the charging had ended successfully.

\ndssparagraph{Prevention of New Charging Sessions}
In addition to disrupting ongoing charging sessions, we also tested the effects of the attack if it commences before a charging session has begun. We started broadcasting the preamble before the vehicle was connected to the charging station. We then tried to start a charging session as described above.
In line with our expectations, the charging did not start and the EVSE displayed an error message. In contrast to the error shown in Figure~\ref{fig:evse_error}, no state-of-charge value was displayed in this case --- as the EVSE had never learned it. 
An adversary could use this variant of the attack to cause a denial-of-service of a certain charging hub for a longer period of time, for example, to blackmail the operator.

\ndssparagraph{Liquid Cooled Cables}
We tested the attack on different high-power DC charging stations with varying maximum charging capabilities ranging from 22~kW up to 350~kW.
Due to the high heat dissipation during the charging process, the charging stations must comply with strict regulations to guarantee a safe operation.
More specifically, for parts of the charging cable that can be grasped during the charging session, IEC 62196-1 defines a maximum temperature of $50^\circ$~C  for metal and $60^\circ$~C  for non-metal parts~\cite{iec62196-1}.
To prevent burn injuries or damage to the cable, it is crucial to operate the equipment within these specified limits.
However, as charging power increases, air cooling becomes inadequate and the use of liquid cooled cables has become necessary~\cite{michelbacher2017enabling}.
We expected that the presence of a liquid jacket in the cable would attenuate the attack signal substantially and make the attack more difficult. 
However, in contrast to our expectations, we could not observe any difficulties interrupting the charging session when the charger was equipped with liquid-cooled charging cables.
We argue that this is due to the coolant only running in between the current-carrying DC wires rather than wrapping around the entire cable, thus excluding the wires (CP and PE) used for the communication.

%% file: 07-Discussion/impact.tex
\section{Impact} \label{sec:impact}

We consider the impact of \textsc{Brokenwire} to be significant. 
In the sections above, we have observed a variety of ways in which the attack could be successfully deployed and found it to be reliable, once suitably configured for the environment. 
Each category of attack within our threat model appears valid. 
For an attacker with a `single vehicle' objective, the attack is achievable at a substantial range and almost irrespective of the target vehicle and charger. 
The only limitation is that the attack could affect other nearby vehicles, so it may be challenging to target only a specific vehicle at range, while leaving others unaffected.
Directional transmission at these frequencies requires a large antenna arrangement, which may limit the attacker if they wish to selectively affect one vehicle from afar.
However, we believe that the ability to miniaturize the equipment and conceal it still provides ample opportunity to mount this attack in a different way.
Having demonstrated the disruption of multiple vehicles at once, we are confident that `fleet denial' and `widespread disruption' attacks are achievable, scaled principally by the effective range of the attack. 
Indeed, in our multiple-vehicle test, we directly emulated the actions of an attacker attempting to render a charging site temporarily useless. 
We consider these as the more serious versions of the attack and so consider further details of their scalability in Section~\ref{sec:parking-bay-estimates} below. 

We also recall that the implications of the attack extend far beyond passenger cars.
We identified a wide range of vehicles that make use of CCS charging directly, along with derivative standards based upon it. 
The Megawatt Charging System (MCS) and SAE J3105 follow the ISO 15118 standard and use PLC for the charging communication~\cite{charin_mcs, sae_j3105}.
While we have not directly tested either, we see no fundamental reason why the attack would not transfer directly to these. In both cases, these derived standards target heavy vehicles with substantial power draws and consider the needs of large, tightly-packed charging parks for fleet use. These factors would make the impact of \textsc{Brokenwire} particularly severe in these cases.

\subsection{Transferability}

Besides the Qualcomm QCA7000 chips that were used by the evaluation boards of our testbed, various other HPGP-compliant modems from other manufacturers exist.
Since we could not access the HPGP modems in our real-world study, we cannot confirm whether any other modem implementations were used in the vehicles and chargers we tested. 
Nevertheless, as the exploited behavior is part of the HPGP standard, we expect that any compliant implementation will be vulnerable, but consider that differences between manufacturers may affect the level of vulnerability on each. 

\subsection{Next-Generation Home Charging}

Our evaluation was heavily focused on DC rapid-charging via the Combined Charging System and Combo~1 and Combo~2 plugs.
However, the new release of ISO 15118-20 standardizes V2G communication and bi-directional charging, and an increasing number of domestic AC chargers with Type-2 plugs are being marketed that offer the same PLC communication link. 
We therefore tested one such new AC charger for domestic use that enables bi-directional charging for the home via ISO 15118.
In accordance with our expectations, the tested charger was also susceptible to the attack and affected in the same way as DC rapid-chargers.
This indicates that the impact of \textsc{Brokenwire} will increase substantially in the near future as home chargers incorporate ISO 15118 and become susceptible. 
Domestic users typically lack the security resources of commercial charge-point operators, in order to defend against attacks.

\subsection{Estimating Affected Parking Bays} \label{sec:parking-bay-estimates}

\begin{figure}[t]
	\centering
	\begin{subfigure}[b]{.99\linewidth}
		\centering
		\includegraphics[width=.99\textwidth]{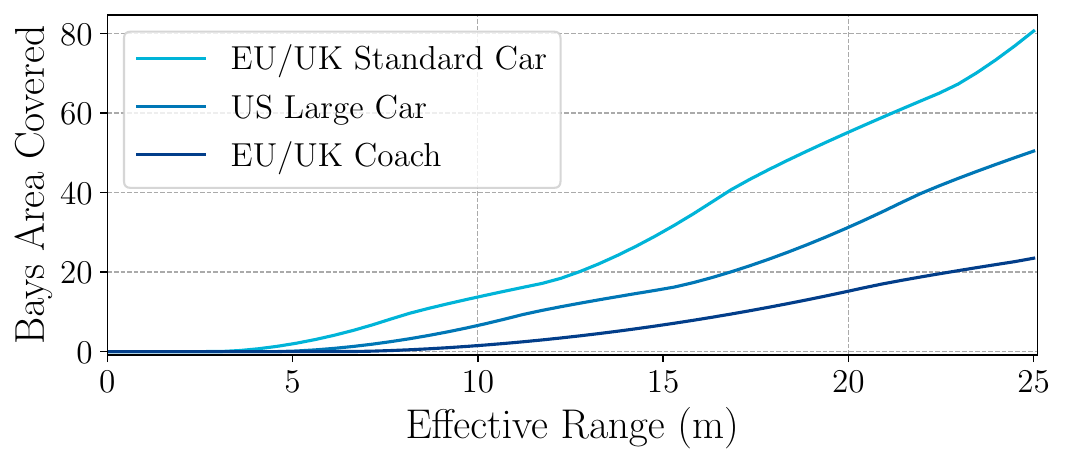}
	\end{subfigure}%
	\caption{Area of parking bays covered by disruption with a given effective range, for three different regulation bay sizes.}
	\label{fig:parking_space_estimate_graph}
\end{figure}

Throughout our experiments in lab settings and real-world environments, we have considered the attack in terms of the distance, or effective range.
However, we also demonstrated the capability for the attack to affect multiple vehicles at once and to work with occluding objects (e.g., cars, barriers) between the attacker and the victim.
As such, we put our results into greater context by estimating the number of parking bays that could be affected by an attack of a given range. 
We stress that effective ranges are subject to variation and therefore so are counts of affected parking bays. 
However, we believe it is helpful to understand the scalability of the attack and that our analysis helps to illustrate other important factors, such as parking lot design.

We built a simulation tool to calculate the coverage of an attack for arbitrary charging park layouts. 
The full details are described in Appendix~\ref{app:parking-coverage-simulation}. 
Figure~\ref{fig:parking_space_estimate_graph} shows the results of this comparison. 
Bays begin to be covered by the disruption signal above 4 m of range. 
For the tightest arrangement of parking bays (small cars, perpendicular parking, double-sided rows) an area equivalent to 80 parking bays would be covered by a 25~m range (with 64 fully overlapped). 
The same range would allow coverage of an area equivalent to 22 coach bays in a perpendicular arrangement (with 16 fully overlapped). 
Stations with overhead charging, such as shown in Figure~\ref{fig:example-bus} are even denser, permitting yet more vehicles to be disrupted.

%% file: 07-Discussion/discussion.tex
\section{Ethical \& Legal Considerations} \label{sec:ethical_considerations}

\ndssparagraph{Safety Measures during the Experiments}
Given the nature of the infrastructure under investigation, we collaborated with a national government department and a local charge-point operator for our evaluation. 
We further took precautions to limit any risk of unintentional effects from our testing. 
We selected only test sites for which no other charging parks were within a reasonable range. 
We only executed the attack when no other vehicles were charging and could immediately abort the experiments if the conditions became uncontrolled. 
Outside our closed laboratory sites, we were limited to a maximum output power of 1~W to ensure our attack signal was compliant with all national transmission regulations.

\ndssparagraph{Responsible Disclosure}
The findings of this paper have been disclosed to different standardization bodies, such as CharIN e.V., a non-profit association founded by car manufacturers and suppliers that leads the standardization of CCS, and Automotive Information Sharing and Analysis Center (Auto-ISAC), as well as to different government entities.
Although our source code is now publicly available, we have embargoed the release of our code during the disclosure period to provide manufacturers with sufficient time to address the vulnerability while making it more difficult for a potential attacker to exploit the vulnerability.
Since the disclosure, the attack has been independently verified and acknowledged by the industry, and CVE-2022-0878 has been published as a result.

%% file: 0X-Countermeasures/countermeasures.tex
\section{Countermeasures}

Preventing electromagnetic interference completely is challenging.
In the following, we discuss potential countermeasures that cannot fully prevent the attack but raise the bar high enough to make the attack too resource-intensive for a would-be attacker.
If the required attack budget increases substantially, the attack becomes potentially unattractive even for more sophisticated adversaries.

\subsection{Electromagnetic Hardening}

Due to the nature of the vulnerability, hardware adjustments are a good first step to improve the security of the affected system. In particular, the two single, unbalanced wires currently used for communication in CCS could either be replaced by a more EMI-resistant alternative (a balanced twisted-pair or coaxial cable) or additionally shielded.

Similar to the protection of any other electronic device, the most straightforward approach to reduce the susceptibility to electromagnetic waves is electromagnetic shielding.
At the same time, wrapping a conductive layer around the CP and PE lines would mitigate data leakage through unintentional electromagnetic emanation~\cite{baker2019}.
While shielding makes the attack more difficult, it does not fully prevent it. 
Instead, it only attenuates the injected signal with the effectiveness largely depending on the thickness of the shielding~\cite{tong2016advanced}.
Since a preamble just above the noise level is sufficient, the adversary can utilize more powerful equipment or reduce the distance to ensure the electromagnetic waves can penetrate the shielding and couple onto the cable.
Therefore, shielding can be seen as an arms race between defender and attacker.

As an alternative to shielding, the cable type can improve the robustness of a communication against external interference.
Instead of using two single, unbalanced wires as used by CCS, a balanced twisted pair cable should harden the communication against injections. 

However hardening is applied, the inclusion of extra wiring or insulation also has practical issues. Charging cables are already heavy, stiff, and sometimes difficult to handle and the addition of additional material would further increase this problem --- making it less usable for drivers while simultaneously more costly for manufacturers. This issue is less pronounced when using twisted pair cabling as it is more flexible. Finally, any modification to the wiring cannot easily be retrofitted onto old cables, so it is necessary to replace the entire charging cable. This is both costly (circa \$1,200 per cable) and time-consuming. However, for future deployments changing the cable can add an additional layer that increases attack difficulty.

\ndssparagraph{Testing Effectiveness}
To test how the required transmission power changes for different cables, we replaced the normal unshielded power cable with a shielded CAT 7. Ethernet cable (SF/FTP), an unshielded CAT 5. Ethernet cable (UTP), and a shielded coaxial cable.
The length of the cables matched exactly the length of the other cable used in previous tests.
The Ethernet cables consisted of four twisted pair sets, but only one was used for the CP and PE.
The twisted pair sets of the CAT 7. cable were individually shielded, in addition to an outer foil and braided shielding. The CAT 5. cable was not shielded at all.
In contrast, the coaxial cable was shielded using a woven copper shield.
For each cable type, we repeated the experiments as described in Section~\ref{sec:attack_evaluation} for distances 1, 2, and 3~m.

\begin{figure}[t]
	\centering
		\subcaptionbox*{} {
		\hspace{-12.5mm}
		\includegraphics[width=0.85\linewidth]{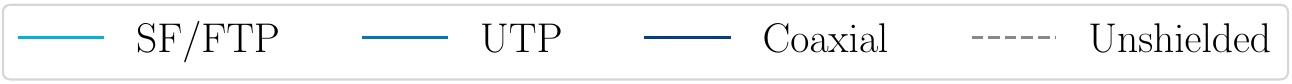}\vspace{-8mm}%
	}
	\begin{subfigure}[b]{.499\linewidth}
		\centering
		\includegraphics[width=.935\textwidth]{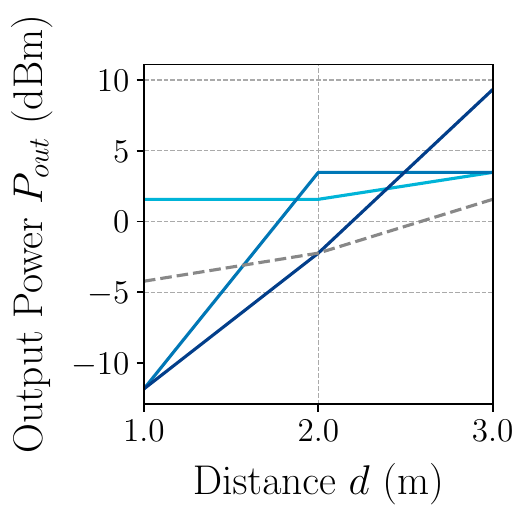}
		\caption{Results in dBm.}
		\label{fig:cable_dBm} 
	\end{subfigure}%
	\begin{subfigure}[b]{.499\linewidth}
		\centering
		\includegraphics[width=.95\textwidth]{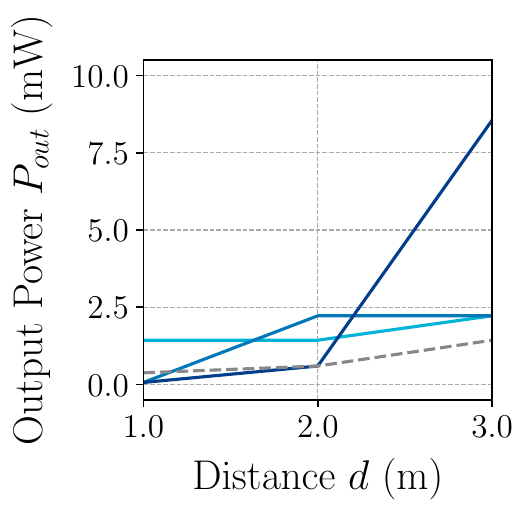}
		\caption{Results in mW.}
		\label{fig:cable_mw} 
	\end{subfigure}%
	\caption{Minimum required transmission power to cause a packet loss of 100\% for different cable types.}
	\label{fig:cable_evaluation}
\end{figure}
The results of the experiments are depicted in Figure~\ref{fig:cable_evaluation}.
To our surprise, the unshielded power cable outperformed the UTP and coaxial cables for close distance.
We argue that this is due to the higher noise resistance of the cables.
As such, not only the attacking signal is coupling less well onto the wire, but also the environmental noise, making it easier to achieve the required SNR of $\geq 2$~dB.
However, with increasing distance, the advantage of better cabling becomes more apparent.
For distances above 2~m, all tested alternatives demonstrated higher resistance against the attack, with the coaxial cable performing particularly well. 
While the testbed with the unshielded power cable was disrupted from 3~m away with around 1.4~mW, the coaxial cable increased the required transmission power to 8.5~mW.
The results indicate that a different cable can substantially increase the power required for a successful attack.
At the same time, the results emphasize that electromagnetic shielding and improved cables on their own are not sufficient to prevent the \textsc{Brokenwire} attack.
Nevertheless, we believe that changing the communication cable in future deployments can easily be implemented and adds an additional layer of defense by raising the bar for the attacker.
\\\\
Since the effectiveness of the attack varied depending on the cable, the results also support our hypothesis that the signal couples mainly onto the charging cable.
However, we would like to note that it is still possible for the signal to couple onto other components, such as the PCB directly or the power lines and transformers to which the charging station is connected.

\subsection{Software Solutions}

Hardware countermeasures are often challenging to deploy.
In particular, the high cost of retrofitting or replacing vulnerable hardware and the limited scalability of the approach make them unattractive.
Therefore, a software-based approach, which can preferably be rolled out via an over-the-air software update, would be optimal.

\ndssparagraph{Optimizing the Channel Access Mechanism}
\textsc{Brokenwire} is inherent to the channel access mechanism implemented in the HPGP standard --- CSMA/CA.
As such, a straightforward approach to fix the vulnerability would be the deactivation of this mechanism.
However, this decision could cause interference and degrade the performance and reliability of the communication.
Therefore, in contrast to entirely deactivating CSMA/CA, a more gentle solution would be to adjust the SNR for which a preamble is detected, seen as valid, and the communication link considered as busy.
Similar to shielding, this approach does not fully prevent the attack; it only requires the adversary to increase their maximum power budget for the attack to be successful.

\ndssparagraph{Enabling Re-Authentication}
The communication is only disrupted as long as the adversary continues to emit the malicious signal.
Once the broadcasting stops and the attacker leaves, the communication link is re-established and the EV and EVSE could continue to communicate. 
However, as discussed in Section~\ref{sec:real_world_evaluation}, none of the tested vehicles automatically continued the charging process once the session was interrupted. 
Instead, the car and charger switched into an error state, forcing the user to repeat the entire authentication process.
We argue that the attack would be less harmful if the vehicle would automatically re-establish the charging session once the communication link is available again.
With the help of the Proximity Pilot (PP), which is part of the CCS Combo~1 and Combo~2 plug and used to detect if the charging cable has been plugged in, it would be possible to monitor if the charging cable has been unplugged.
If the circuit has not been interrupted, the charger is still connected to the same vehicle, which has previously been authenticated successfully by the user, making a manual re-authentication by the user unnecessary.
With the introduction of Plug \& Charge, the re-establishment of the charging session is even easier.
The car has all the necessary information to start charging without the need for the user to authenticate.
While this countermeasure does not solve the vulnerability, it substantially reduces the impact of one-off attacks, such as the drive-by attack.

\ndssparagraph{Internal Counter}
Performance and error metrics are typically tracked by modems internally and sometimes exposed to higher layers in the host. 
These could be used to detect the abnormal situation that \textsc{Brokenwire} creates. 
For example, during an attack, the number of new packet detections increases to an abnormally high level --- beyond the maximum that could ever occur in normal operation. 
The number of invalid packets also increases, as only the preamble is transmitted and no valid data follow. 
Detecting an impossible packet detection rate and a packet error rate approaching 1.0 would indicate the presence of the attack. 
This approach does not prevent the attack, but could be use to drive other, reactive countermeasures or reported to the driver and charge-point operator for further action.

\subsection{Increasing the Noise Floor}

While the aforementioned solution would be ideal, it might not be possible to update the firmware of the PLC modem.
Therefore, we propose a simple and easy-to-deploy defense technique that is highly effective and only costs a fraction of what it would cost to replace the entire charging cable.
As mentioned in Section~\ref{sec:evaluation_of_noise}, we observed that the PLC communication is surprisingly robust to Gaussian white noise.
Even a very powerful noise signal was not sufficient to degrade the communication link quality.
Our countermeasure exploits this fact to increase the difficulty for the adversary to execute a successful attack.
As a reminder, the \textsc{Brokenwire} attack relies on the injection of a preamble with an SNR $> 2$~dB.
This means that if the background noise on the wire increases, the adversary requires more power to keep the SNR of the injected preamble above this threshold.
We propose to add a small noise source to the PLC modem to intentionally increase the noise on the communication lines.
Usually, noise needs to be prevented to guarantee a robust communication, however, in this scenario, noise contributes to the defense against the \textsc{Brokenwire} attack.
We admit that this countermeasure might not be optimal, yet, we argue that the advantages, cheap and easy to deploy, outweigh the disadvantages.

\ndssparagraph{Method}
To test the effectiveness of the countermeasure, we repeated the experiments as described in Section~\ref{sec:attack_evaluation}.
We only modified the experimental setup slightly to enable the injection of noise. 
More precisely, we connected a Picoscope 5244D as a noise source directly via the SMA connector provided by the PLC evaluation boards to the communication lines.
We then started a UDP IPerf session and re-ran the attack for different distances while injecting Gaussian white noise with varying amplitude onto the wires.

\ndssparagraph{Results}
\begin{figure}[t]
	\centering
		\subcaptionbox*{} {
		\hspace{-7.5mm}
		\includegraphics[width=0.70\linewidth]{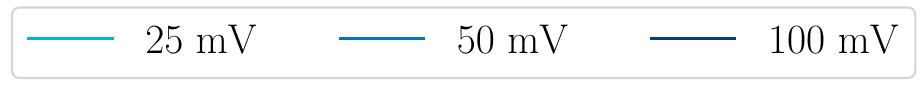}\vspace{-8mm}%
	}
	\begin{subfigure}[b]{.499\linewidth}
		\centering
		\includegraphics[width=.935\textwidth]{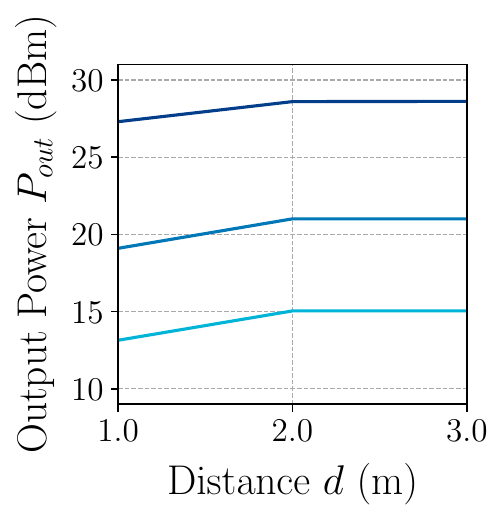}
		\caption{Results in dBm.}
		\label{fig:defense_dbm} 
	\end{subfigure}%
	\begin{subfigure}[b]{.499\linewidth}
		\centering
		\includegraphics[width=.95\textwidth]{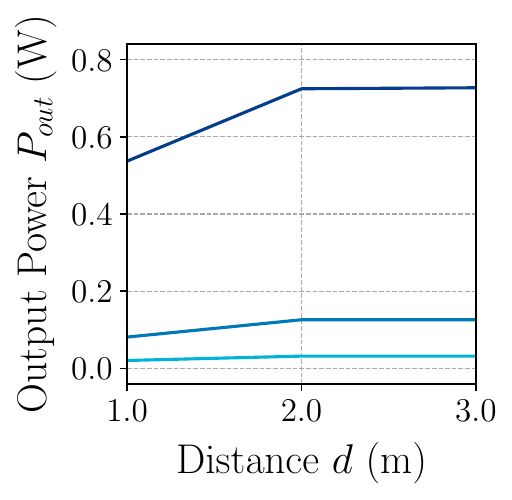}
		\caption{Results in W.}
		\label{fig:defense_w} 
	\end{subfigure}%
	\caption{Minimum required transmission power to cause a packet loss of 100\%. With increasing background noise, the attack becomes progressively more difficult.}
	\label{fig:countermeasure_evaluation}
\end{figure}
We found the countermeasure to be effective even for close attack distances.
In Figure~\ref{fig:countermeasure_evaluation}, the minimum required transmission power to completely disrupt the communication for different distances under various noise levels is depicted.
The results show that even for a distance of 1~m, noise with an amplitude of only 25~mV was adequate to increase the required transmission power for a successful attack by a factor of 50. 
For a packet loss of 100\%, a transmission power of 20.6~mW was required compared to 0.37~mW without the countermeasure in place. 
Given the maximum output power of around 1~W of our attack setup, noise with an amplitude of 25~mV was sufficient to prevent the attack from a distance of 10~m.
As such, for a fixed power budget, the maximum attack range is considerably reduced, making wardriving or attacks at larger charging parks far more challenging, expensive and potentially unattractive.
Surprisingly, the power required to disrupt the testbed from 2 and 3~m away was almost identical. 
As mentioned earlier, we consider this to be a side-effect of the uncontrollable background noise in our lab environment.
In general, the results underline our hypothesis that a higher noise floor makes the attack inevitably more difficult.
This is also in agreement with our assumptions drawn from our real-world evaluation discussed in Section~\ref{sec:real_world_evaluation}, where we observed that the required output power varies between charging station locations.

%% file: 0X-RelatedWork/relatedwork.tex
\section{Related Work}

Disrupting wireless communication has been well studied~\cite{mitch2011signal, bayraktaroglu2013performance, poisel2011, xu2005feasibility, wilhelm2011short}.
Nevertheless, simple jamming that aims to decrease the signal-to-noise ratio by emitting noise in the spectrum of the communication channel is ineffective and resource-intensive.
As such, more intelligent jamming strategies that exploit design weaknesses in the targeted protocol, for example, IEEE 802.11, LTE, or 5G, have been demonstrated~\cite{clancy2011efficient, rahbari2015swift, la2016physical, arjoune2020smart}. 
In addition, denial-of-service attacks utilizing vulnerabilities at the Medium Access Control (MAC) layer have proven effective~\cite{negi2003jamming}.
Exploiting the CSMA/CA mechanism in IEEE 802.11 networks has been discussed by different papers~\cite{toledo2008robust, zhangpreamble, kyasanur2005selfish, pelechrinis2010denial}, with~\cite{zhangpreamble} being the closest to our work.
The authors demonstrated that the throughput of a WiFi channel can be substantially reduced by injecting fake preambles that cause the target node to back off and stop communicating.

It is well known that PLC is susceptible to electromagnetic interference (EMI)~\cite{nateghi2021susceptibility, soriano2017feasibility, lampe2016}.
At the same time, PLC tends to cause electromagnetic emanation, even so strong that it can interfere with amateur radio~\cite{lampe2016}. 
The work by~\cite{baker2019} demonstrated that this also applies to the Combined Charging System.
A similar attack, although wired rather than wireless, has been demonstrated in~\cite{dudek2019}.
However, to the best of our knowledge, no research has been conducted that evaluates the real impact of EMI against the CCS charging communication.
The authors of~\cite{bao2018threat} mentioned only briefly the possibility of a denial-of-service attack against the charging communication by emitting noise in its spectrum. 
And the researchers in~\cite{dayanikli2020electromagnetic} focused on the electromagnetic susceptibility of the voltage and current sensors in charging stations.

%% file: 0X-Conclusion/conclusion.tex
\section{Conclusions}

In this paper, we presented \textsc{Brokenwire}, a wireless attack against the Combined Charging System (CCS), the most widely used DC rapid charging standard for electric vehicles in North America and Europe.
We investigated the effects of the \textsc{Brokenwire} attack in a controlled laboratory environment and an extensive real-world evaluation, including eight EVs and 20 charging stations.
We demonstrated that the attack can be conducted with only off-the-shelf equipment and with little knowledge, making the entry barrier for an attacker low.
Based on the insights that we gained from our evaluation, we proposed, examined, and compared different mitigation strategies.
Our results suggest that the use of PLC for charging communication is a serious design flaw that leaves millions of vehicles, some of which belong to critical infrastructure, vulnerable.

\section*{Acknowledgments}
We are grateful for the support from Armasuisse S+T and EWZ (Elektrizitätswerk der Stadt Zürich).
We would also like to thank Daniel and Peter Köhler for providing access to their vehicles and supporting us during some of the experiments.
Sebastian was supported by the Hans-Böckler Foundation and the Engineering and Physical Sciences Research Council (EPSRC).

\section*{Availability}
Our evaluation source code is available at \url{https://github.com/ssloxford/brokenwire}.
To facilitate deployment and make it easier for the community to reproduce our results, the entire project is dockerized.

%% file: 0X-Appendices/appendices.tex
\ndssparagraph{Comparison to Friis Equation}
\label{app:friis-comparison-details}

In order to make estimates of transmission power requirements, we first measured the ambient noise level on the cable, finding it to be 2.45 mV. The HPGP standard states a conformance threshold for detection of preambles (Section 3.8.4.2~\cite{alliance2013homeplug}):

\begin{quote}
The receiver shall be able to detect the presence of Preamble Symbols within a Slot Time with a miss rate not exceeding 1\% using the standard North American mask [...] \\
When the desired Preamble Symbol waveform present at the receiver has a signal power
of -35 dBm and is corrupted by Gaussian noise producing a total SNR of 2 dB at the
receiver terminal.
\end{quote}

providing an equation as:

\begin{equation}
    SNR_{dB} = 10 \cdot log_{10} ( \frac{mean(x(t)^2)}{mean(n(t)^2)} \cdot \frac{BW_{Noise}}{BW_{Signal}} )
\end{equation}

with $x(t)$ as the time-domain signal, $n(t)$ as the noise signal and $BW_{Noise}$, $BW_{Signal}$ as the bandwidths of noise and signal measured in subcarriers. 

\noindent Based on this, we re-arrange as follows, simplifying the $mean()$ of time-domain signals as our values are already averages:

\begin{equation}
    \hat{x}^2 = 10^{\frac{SNR_{dB}}{10}} \cdot (\hat{n}^2 \cdot BW_{Signal}) \cdot \frac{1}{BW_{Noise}}
\end{equation}

and divide for specified cable impedance of 100 $\Omega$~\cite{devolo_devboards}. 

\noindent Setting the value of $n$ to 2.45 mV and taking $SNR_{dB}$ as 2, per the standard, we can compute the minimum power of a preamble signal that must be detected in order to comply. Taking a common formulation of the Friis equation~\cite{friis1946note}:

\begin{equation}
    P_R = \frac{P_T G_T G_R \lambda^2}{(4 \pi R)^2}
\end{equation}

we rearrange as:

\begin{equation}
    P_T = \frac{P_R (4 \pi R)^2}{G_T G_R \lambda^2}
\end{equation}

\noindent We model the antennas as dipole and monopole and estimate their gains from typical values ($G_T$ = 2 dBi, $G_R$ = 4 dBi) and apply this equation to compute the expected transmission power $P_T$ for a signal, assuming free-space path loss and a distance of $R$. The results are shown in Figure~\ref{fig:disruption_results_vs_maths} in the main text.

\ndssparagraph{Parking Bay Coverage Simulation}
\label{app:parking-coverage-simulation}

\begin{figure}[t]
	\centering
	\centering
	\includegraphics[width=1\linewidth]{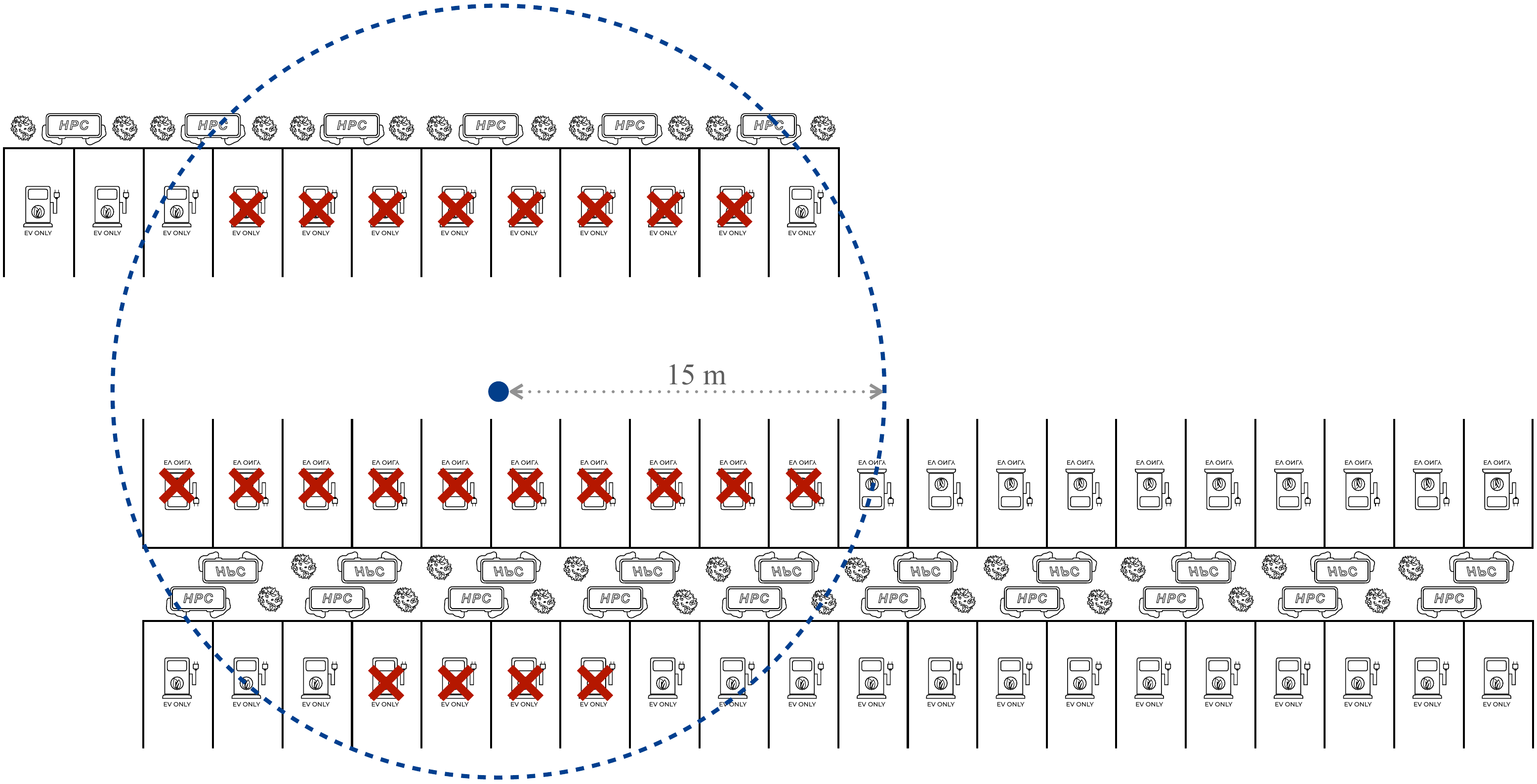}
	\caption{Layout of the charging hub shown in Figure~\ref{fig:hpc_hub} demonstrating that with 15~m attack range, an adversary can disrupt the charging of roughly 22 vehicles simultaneously.}
	\label{fig:parking_space_estimate_example}
\end{figure}

In order to help assess the scalability of \textsc{Brokenwire}, we created a simulation tool to help estimate how the effective range of an attack covers bays in parking lots. 

In Figure~\ref{fig:parking_space_estimate_example}, we illustrate the layout of the charging hub pictured in Figure~\ref{fig:hpc_hub} and show how a `drive-by' attack with an effective range of 15~m could be expected to affect approximately 22 vehicles at once. 

However, the design of parking lots varies substantially, not only in terms of regulations on sizing and angle, but also in layout, which is subject to site constraints and design choices by the land developers. In Figure~\ref{fig:parking_space_estimate_example} there is a central, double-sided row (i.e., bays on both sides of a sidewalk) and another single-sided row, offset in the top-left. 

Given this variation, our simulation accepts as input an arbitrary description of the layout of a parking lot, along with an attacker location and effective range. The simulation then estimates the number of parking bays that could be affected by an attack of that size. Bays can be defined with any size, shape and orientation, allowing the simulation to consider bays suitable for any type of vehicle. An attacker can be situated at any point within the layout --- including outside the parking lot. They are modeled as a point-source with a uniform, circular radiation pattern. The simulation computes the total area of parking bays covered by an attack of a given range. Although we have shown effectiveness of the attack between floors, we implement only the 2-dimensional case here. We make the simulation tool available as part of our open source code. 

For the analysis in the main text, we consider an example parking lot arrangement. This does not follow a specific real-world case, as in Figure~\ref{fig:parking_space_estimate_example}, but a generally common design. While there are site-specific variations, parking bays are typically arranged in regular patterns, broken only by access roads and site boundaries. We use a double-sided arrangement in repeated rows, with chargers between rows, perpendicular parking and a range of bay sizes and spacings from regulations around the world~\cite{parkingbays-ise,parkingbays-reid}. EU/UK `normal' bays are 2.4~m $\times$ 4.8~m with access road width of 6~m. US `large' bays are 3.2~m $\times$ 6.5~m with 9.2~m roads. EU/UK coach bays are 3.5~m $\times$ 14~m with 13~m roads. 

The parking lot is scaled to be sufficiently large for any effective range selected. Our selected layout represents the most dense arrangement used in real parking lots --- such as in Figure~\ref{fig:parking_space_estimate_example}. As such our results give an upper bound on the scale of the attack, yet still a realistic one. 

We selected an attacker position that maximizes the coverage of parking bays: centrally-positioned in the parking lot on an access road. We simulate for distances up to 25~m, noting that while we achieved success at nearly double this range, charging parks with more than 50 fast-charging bays are all but non-existent at time of writing.